\documentstyle[12pt]{article}

\font\ten=cmbx10 at 12pt

\renewcommand{\thefootnote}{\fnsymbol{footnote}}

\newcounter{saveeqn}
 \newcommand{\alpheqn}{\setcounter{saveeqn}{\value{equation}}%
  \stepcounter{saveeqn}\setcounter{equation}{0}%
   \renewcommand{\theequation}{\mbox{\arabic{saveeqn}-\alph{equation}}}}
    \newcommand{\reseteqn}{\setcounter{equation}{\value{saveeqn}}%
     \renewcommand{\theequation}{\arabic{equation}}}

\newenvironment{goodlist}[1]%
{\begin{list}{}{\settowidth{\labelwidth}{#1}
  \setlength{\leftmargin}{\labelwidth}
  \addtolength{\leftmargin}{\labelsep}
  \setlength{\parsep}{1ex plus0.7ex minus0.7ex}
  \setlength{\itemsep}{0.8ex}
  }}{\end{list}}

\def\build#1_#2^#3{\mathrel{\mathop{\kern 0pt#1}\limits_{#2}^{#3}}}

\def\th{$^{\mbox{\scriptsize th}}$}

\begin{document}

\begin{titlepage}

\begin{center}

{\ten Centre de Physique Th\'eorique\footnote{Unit\'e Propre de
Recherche 7061} - CNRS - Luminy, Case 907}

{\ten F-13288 Marseille Cedex 9 - France }

\vspace{1 cm}

{\large\bf PHENOMENOLOGICAL APPROACH TO UNPOLARIZED AND POLARIZED PARTON
DISTRIBUTIONS AND EXPERIMENTAL TESTS }

\vspace{0.3 cm}
\setcounter{footnote}{0}
\renewcommand{\thefootnote}{\arabic{footnote}}

{\bf Claude BOURRELY and Jacques SOFFER}

\vspace{1 cm}

{\bf Abstract}

\end{center}

We recall our recent description of quark parton densities of the proton
at $Q^2=4 GeV^2$ in terms of Fermi-Dirac distributions parametrized
with very few free parameters. We have also proposed some simple
assumptions to relate unpolarized and polarized quark parton densities
which lead to a fair description of the spin-dependent structure
functions $xg^p_1(x,Q^2)$ and $xg^n_1(x,Q^2)$ at low $Q^2$. We will
show the predictions we obtain after a straightforward DGLAP $Q^2$
evolution and comparison in a much broader $x$ and $Q^2$ range, with
several recent and accurate deep-inelastic scattering data. In
particular, we will see that we get an excellent agreement with the
sharp rise of $F^{ep}_2(x,Q^2)$ for small $x$, recently observed at HERA.
Finally, we give several predictions for lepton pair and gauge boson
production in $pp$ and $pn$ collisions at high energies which will be
tested in the future at RHIC.

\vspace{1 cm}


\bigskip

\noindent Number of figures : 31

\bigskip

\noindent February 1995

\noindent CPT-95/P.3160

\bigskip

\noindent anonymous ftp or gopher: cpt.univ-mrs.fr

\end{titlepage}

\section{Introduction}
Deep-inelastic lepton-hadron scattering is the basic source of our
knowledge on quark parton densities of the proton and an enormous
amount of data has been accumulated over the last twenty years or so.
It allows to test rather accurately our picture of the nucleon
structure at short distances. With the advent of HERA, a new
kinematical range is now accessible in $x$ down to $10^{-4}$ and in
$Q^2$ up to $10^4 GeV^2$, which will be extremely useful for the
physics analysis at future hadron colliders. Besides the determination,
for each flavor $u,d,s,$ etc..., of the valence quarks which dominate,
say for $x\geq 0.1$, and of the sea quarks (or antiquarks)
which dominate in the small $x$ region, it is very important to extract
the correct gluon distribution. This is because, first it also
prevails at low $x$ and second, it plays a crucial role in the $Q^2$
evolution for testing perturbative QCD.

Let us now briefly review the main points of our approach. Since quarks
carry a spin$-1/2$, it is natural to consider that the basic
distributions are $xq^{\pm}_i(x,Q^2)$ corresponding to a quark of
flavor $i$ and helicity parallel or antiparallel to the proton
helicity. Recently we have proposed \cite{BS} a simple description of
these quark densities in terms of Fermi-Dirac distributions with very
few free parameters, which were determined from the data at
$Q^2=Q^2_0=3GeV^2$. This statistical physical picture for the nucleon
structure functions is largely motivated by the importance of the Pauli
exclusion principle, which can be advocated to explain several
experimental features, as discussed in ref.\cite{BS}. The fact that
there is also some experimental evidence for the existence of simple
relations between unpolarized and polarized quark parton densities has
allowed us to reduce to {\it four}, the total number of independent
distributions, i.e. two for valence quarks $xu^{\pm}_{val}(x,Q^2_0)$
and two for sea quarks (or antiquarks) $xu^{\pm}_{sea}(x,Q^2_0)$. In
this approach, the corresponding four $d$ quark densities
$xd^{\pm}_{val}(x,Q^2_0)$ and $xd^{\pm}_{sea}(x,Q^2_0)$ are related to
the four $u$ quark densities mentioned above and similarly for the $s$
quark (or antiquark) densities. Concerning the gluon distribution
$xG(x,Q^2)$, for the sake of consistency, at $Q^2=Q^2_0$ we have used a
Bose-Einstein expression. As we will see, on the one hand, it is fully
consistent with our present knowledge of $xG(x,Q^2)$ extracted from
deep-inelastic scattering data and, on the other hand, it leads to the
correct $Q^2$ evolution of the structure functions. Actually, one of
the main purpose of this paper is to show that, starting from our
simple scheme, which is in very good agreement with all spin-average
and spin-dependent structure functions for $Q^2$ near $Q^2=Q^2_0$, one
gets, after a standard QCD $Q^2$ evolution, {\it predictions} fairly
consistent with the data, in a much broader $x$ and $Q^2$ range.

The paper is organized as follows. In section 2, we briefly review and
update our parametrization of the quark (antiquark) parton and gluon
densities at $Q^2=Q^2_0$ and show their $Q^2$ evolution. In section 3,
the predictions we obtain for the spin-average structure functions
$F^{\ell p}_2(x,Q^2)$ with $\ell=e,\mu$ and $xF^{\nu N}_3(x,Q^2)$ are
compared with various deep-inelastic scattering data, including the
recent results from HERA. We also give our predictions for the $W$
production cross section at the Tevatron and for the asymmetries in
Drell-Yan lepton pair production and $W$ production in $pp$ and $pn$
collisions, which are sensitive to the flavor asymmetry of light sea
quarks in the proton. Section 4 is devoted to the discussion of quark
and gluon helicity distributions at $Q^2=Q^2_0$ and the comparison of
our predictions with the most recent data on the spin-dependent
structure functions $xg^p_1(x,Q^2)$ for proton, $xg^n_1(x,Q^2)$
for neutron and $xg^d_1(x,Q^2)$ for
deuteron. We also discuss their $Q^2$ evolution and give our
predictions for single and double helicity asymmetries for $pp$
collisions at RHIC. Finally, in section 5 we consider the {\it
transversity} distribution of quarks (antiquarks) denoted by
$h^q_1(x)$, which can be extracted from Drell-Yan lepton pair and $Z$
production in $pp$ collisions with both proton beams transversely
polarized. We give our concluding remarks in section 6.

\section{Quark (antiquark) parton and gluon distributions} Let us first
consider the distributions at
$Q^2=Q^2_0=3GeV^2$. For valence quarks we have two basic densities
$xu^{\pm}_{val}(x,Q^2_0)$ which are expressed in terms of Fermi-Dirac
distributions of the form
\begin{equation}\label{xp}
xp(x,Q^2_0)=\frac {a_px^{b_p}}{exp[(x-\widetilde{x_p})/\bar x]+1}
\end{equation}
where $\widetilde{x_p}$ plays the role of the ''thermodynamical
potential'', which is a constant for each quark species and $\bar x$ is
the ''temperature'' which is the same for quarks, antiquarks and
gluons. This {\it universal constant} is independent of the parton
helicity. We use a fitting procedure slightly different from that of
ref.\cite{BS} which is based on the most accurate neutrino data from CCFR
\cite{Q,F} for $xF_3^{\nu N}(x,Q^2)$, which has pure valence
contributions, and also gives the antiquark distribution at $Q^2=3
GeV^2$ (see below). Of course we will check the good agreement with the
most recent NMC data\cite{AR} for $F^p_2(x,Q^2)$ and $F^n_2(x,Q^2)$ at
$Q^2=4 GeV^2$. It yields the following universal temperature and four
free parameters entering in eq.(\ref{xp})
\begin{equation}\label{x}
\bar x=0.092,~ b_+=0.354,~b_-=0.738,~\widetilde{x_+}=0.510,
{}~\widetilde{x_-}=0.231\ .
\end{equation}

Clearly one has
\begin{equation}\label{xu}
xu_{val}(x,Q^2_0)=xu^+_{val}(x,Q^2_0)+xu^-_{val}(x,Q^2_0)\ ,
\end{equation}
and, as we argued in ref.\cite{BS}, we take
\begin{equation}\label{xd}
xd_{val}(x,Q^2_0)=2xu^-_{val}(x,Q^2_0)\ .
\end{equation}
$a_{\pm}$ are not free parameters, but two normalization constants for
the valence quarks in the proton. We note that the values of these
parameters (see eq.(\ref{x})) are different from those of ref.\cite{BS}
and correspond to a better fit of $xF_3^{\nu N}(x,Q^2_0)$ at large $x$,
as shown in Fig.1a. In particular we now find that the temperature $\bar
x$ is slightly smaller than what we found in ref.\cite{BS}. We also
note that the potentials $\widetilde{x_+}$ and $\widetilde{x_-}$ for
the two different helicity states have numerical values in the ratio
$\sqrt 5$ which is smaller than what we found in our previous analysis.
As expected $\widetilde{x_+} > \widetilde{x_-}$, so $u^+_{val}$
dominates over $u^-_{val}$.

We now turn to antiquarks (or sea quarks) and for $x\bar
u^{\pm}(x,Q^2_0)$ we use the same expression eq.(\ref{xp}), but in this
case the potential has a smooth $x$ dependence. This might reflect the
fact that in this statistical description of the proton, we must
consider the existence of two phases : a gaz corresponding to valence
quarks, which dominates at large $x$, with a constant potential and a
liquid corresponding to sea quarks (or antiquarks), which prevails at
small $x$, with a potential slowly varying in $x$, that we take linear
in $\sqrt x$. In addition, we expect quarks and antiquarks to have
opposite potentials, consequently the gluon which produces $q\bar q$
pairs has a zero potential (see below). Moreover since in the process
$G\to q_{sea}+\bar q$, $q_{sea}$ and $\bar q$ have opposite helicities,
the potentials for $u^+_{sea}$ (or $\bar u^+$) and $\bar u^-$ (or $
u^-_{sea}$) will be opposite. So we take
\begin{equation}\label{xtilde}
\widetilde
x_{\bar u^+}=-\widetilde x_{\bar u^-}=x_0+x_1\sqrt x\ .
\end{equation}
For the $\bar d$ distributions, we assume no polarization at $Q^2_0$
 so we take
\begin{equation}\label{xbard}
x\bar d^+(x,Q^2_0)=x\bar d^-(x,Q^2_0)=x\bar u^-(x,Q^2_0)\ ,
\end{equation}
in accordance to eq.(\ref{xd}). Since when $x\to 0$, from Pomeron
universality, one expects $x\bar u(x,Q^2_0)=x\bar d(x,Q^2_0)\not = 0$,
$\bar a_-$ is not a free parameter and will be fixed by this
constraint. This implies in addition $\bar b_+=\bar b_-$. For the
strange quark (or antiquark) distributions, we also assume they are
unpolarized and we take, in agreement with neutrino deep-inelastic
scattering data,
\begin{equation}\label{xs}
xs(x,Q^2_0)=x\bar s(x,Q^2_0)=\frac{1}{4}\left(x\bar u(x,Q^2_0)+ x\bar
d(x,Q^2_0)\right)\ .
\end{equation}
Therefore the antiquarks depend on {\it four} free parameters which are
different from those determined in ref.\cite{BS}, that is
\begin{equation}\label{a}
\bar a_+=0.0185,\ \bar b_+=\bar b_-=-0.340,\
x_0=0.219\quad\hbox{and}\quad x_1=-0.406\ .
\end{equation}

We show in Fig.1b the result of our fit for $x\bar q(x)$ at $Q^2=3
GeV^2$. The NMC data for $F^p_2$ and $F_2^n$ which involve both valence
quarks and sea quarks have also been used to test the correct
determination of our four basic quark (antiquark) densities and the
results are shown in Figs.2a,b.

For comparison, we have also given in Figs.1 and 2 the curves obtained
by using the MRS(A) parametrization at $Q^2=4GeV^2$, taken from
ref.\cite{MRS}.

 Finally concerning the gluon distribution, for the sake
of consistency, we use a Bose-Einstein expression given by
\begin{equation}\label{xg}
xG(x,Q^2_0)=\frac{a_Gx^{b_G}}{e^{x/\bar x}-1}
\end{equation}
with a vanishing potential and the same temperature $\bar x$ as we
discussed above. It is also reasonable to assume that for very small
$x$, $xG(x,Q^2_0)$ has the same dependence as $x\bar q(x,Q^2_0)$, so we
will take $b_G=1+\bar b$, where $\bar b=\bar b_{\pm}$ is given in
eq.(\ref{a}). So except for the overall normalization $a_G$,
$xG(x,Q^2_0)$ has no free parameter. From the momentum sum rule, we
find $a_G=13.146$. For the sake of completeness, we also need to
specify the gluon polarization for which there is no data at all, but
we will make some simple speculations in section~4.

To summarize, this statistical approach of the nucleon allows the
construction of {\it all} quark, antiquark and gluon distributions in
terms of simple expressions which depend on {\it nine} free parameters.
We gave some comprehensive arguments about the physical meaning of
these parameters for which we don't have yet a full understanding. This
small number of free parameters has been determined from deep-inelastic
scattering data \cite{Q,F,AR} at low $Q^2$. Then, by means of a
straightforward DGLAP \cite{GL} $Q^2$ evolution, we will compare our
predictions with the existing data in a broad kinematical range of $x$
and $Q^2$, in order to test scaling violations and the dynamics implied
by perturbative QCD. This approach contrasts with other methods
presented in the literature \cite{MRS,CTEQ} where one performs a global
analysis with parton densities expressed in terms of a large number of
free parameters, of the order of twenty, which are fixed by fitting
several hundreds of data points. Notice that they are only dealing with
spin-average distributions and it is necessary to introduce a new set
of parameters to describe spin-dependent structure functions \cite{GS}.
Before closing this section, we would like to show the $x$-shapes of
various unpolarized parton distributions at $Q^2=Q^2_0=3 GeV^2$ and
$Q^2=20 GeV^2$, after a standard $Q^2$ evolution. As already mentioned
in ref.\cite{BS}, we have used a numerical solution \cite{KS} of the
DGLAP equations. We display in Fig.3 the $u$ and $d$ (valence $+$ sea)
quark distributions and also the corresponding antiquark distributions
at two different values of $Q^2$. The gluon density $xG(x,Q^2)$ is
shown in Fig.4. In the next section we will compare the results of our
calculations, using these parton densities, with several pieces of
existing experimental data and we will also give our predictions for
future measurements.

\section{Experimental tests for unpolarized parton densities}
Since all parton densities have been determined, in order to test our
approach, we can now proceed and calculate several physical quantities
measured either in deep-inelastic scattering or in hadronic collisions.
We will first study various structure functions in deep-inelastic
scattering and then turn to hadronic processes with a special emphasize
for testing the flavour asymmetry of the light sea quarks, i.e. $\bar
u\not =\bar d$.

\subsection{Deep-inelastic scattering}
We first consider the high statistics $\nu N$ deep-inelastic scattering
CCFR data \cite{Q} from which one extracts $xF_3^{\nu N}(x,Q^2)$. As
noticed above, this structure function gives a precise measurement of
the isolated valence quark contributions and we show in Fig.5 the
results of our calculations. These neutrino data were obtained on a
iron target, but we don't think we can treat properly the nuclear
effects. Therefore we have not tried to make any heavy nuclear target
corrections which are known to be less important for $xF_3^{\nu N}$
than for $xF_2^{\nu N}$ which contains the sea quark distributions.
However in order to get the remarkable agreement displayed in Fig.5 we
had to shift up by $4\%$ our theoretical predictions.

Next we turn to $\mu p$ and $ep$ deep-inelastic scattering for which
several experiments have yielded a large number of data points on the
structure function $F_2^{\ell p}(x,Q^2)$. First we will analyze the
fixed target measurements which cover the limited kinematical region
$0.0125\leq x\leq 0.75$ and $3\leq Q^2\leq 230 GeV^2$, one obtains by
combining the NMC data \cite{AM} and the BCDMS data \cite{B}. Our
predictions are now compared with the data in Figs.6a,b but in this
case we need to shift up by $12\%$ the theoretical curves. The
description of the data is rather impressive except at $x=0.75$ for
$Q^2<10^2 GeV^2$ where the BCDMS data lie above the prediction. This is
surprizing to us, because this kinematic region is dominated by valence
quark contributions which are in perfect agreement with the measurement
of $xF_3^{\nu N}$ at large $x$, as we have seen in Fig.5.

We now turn to the very recent measurements of $F_2^{ep}$ at Hera from
the Zeus Collaboration \cite{D} and the H$1$ Collaboration \cite{BR}.
It is important to note that these data are essentially in the low $x$
region which is dominated by the sea quark densities. The behaviour of
$F_2^{ep}$ is therefore mainly constrained by the parameters given in
eq.(\ref{a}) and specially by $\bar b=-0.340$, which was fixed by the
steep behaviour of $x\bar q(x)$ at low $Q^2$, as shown in Fig.1b. So
this rise of the Hera data in the small $x$ region was predictable from
the CCFR measurement of $x\bar q(x)$ and no new parameter is requiered.
This is not what was claimed in ref.\cite{MRS} where $\lambda=-0.3$ was
introduced at posteriori. Our predictions, with {\it no overall
normalization factor}, are compared with the data in Figs.7a, b and c and
the agreement is absolutely remarkable. From these new measurements of
$F_2^{ep}$, by analyzing the scaling violations, it is possible to
improve our knowledge on the gluon density in the low
$x$ region. Going back to Fig.4, we see that $xG(x,Q^2)$ increases very
rapidly, and as we expect, it is fairly consistent with the
distribution extracted at $Q^2=20 GeV^2$ from the Zeus Collaboration
\cite{Dal}.

So far we have tested our parton densities in a kinematical region with
$Q^2>Q^2_0=3 GeV^2$ and we have seen that one gets a very satisfactory
description of all deep-inelastic structure functions. Clearly this
confirms the success of the DGLAP $Q^2$ evolution for values above our
starting point at $Q^2=Q^2_0$. Without trying to make a systematic
study of the low $Q^2$, low $x$ region in electroproduction, we would
like to show what we obtain for $Q^2<Q^2_0$, corresponding also to
rather small $x$ values. In this region there are already some data
{}from NMC \cite{AM}, also earlier from SLAC \cite{W} and some
preliminary results from the FNAL muon scattering E665 experiment
\cite{K}. Just for illustration, we compare in Figs.8a and b our
predictions with the NMC and SLAC data for $Q^2\geq 1 GeV^2$ and our
curves for $x\leq 0.007$ are given to be confronted with the final E665
data. For $Q^2< 1 GeV^2$ the understanding of the electroproduction
structure functions certainly lies outside a perturbative QCD framework
and involves other dynamical concepts \cite{BK}.

\subsection{Hadronic processes}
As we discussed in ref.\cite{BS}, one of the main features of our
approach is the flavor symmetry breaking of the light sea quarks which
was first recognized in ref.\cite{PRS} following an earlier NMC
measurement \cite{Aal}. Here we will stress again the importance of
this fact, we will indicate how one can confirm it and we will see
where to expect $\bar u(x,Q^2)\not = \bar d(x,Q^2)$, for different $x$
and $Q^2$ values.

Let us first consider the $W$ production cross section which provides
an important test of the quark densities determined from deep-inelastic
scattering. For illustration we show in Fig.9 the rapidity distribution
for $\bar p p\to W^++X$ at Tevatron energy, which has been computed in
the Drell-Yan picture, dominated by the product $u(x,M^2_W)d(x,M^2_W)$
with $x\sim 0.05$. We find that the maximum value of $d\sigma/dy_W$ is
around $2.6 nb$ which is somehow smaller compared to other predictions
presented in ref.\cite{MRS}. For the total $W$ cross section we find
$\sigma_W=19 nb$ which is compatible with recent experimental values
{}from CDF and $D_0$ at FNAL \cite{ABE}. Next we turn to hadronic
processes dominated by quark-antiquark annihilation, in order to study
further the flavor symmetry breaking in the light sea quarks of the
proton. One possibility, as noticed in ref.\cite{ES}, is to compare
dilepton production in $pp$ and $pn$ collisions by means of the
Drell-Yan asymmetry
\begin{equation}\label{ADY}
A_{DY}=\frac{d\sigma_{pp}/dy-d\sigma_{pn}/dy}
{d\sigma_{pp}/dy+d\sigma_{pn}/dy}\ .
\end{equation}
These cross sections can be easily written in terms of $q(x_a,M^2)$ and
$\bar q(x_b,M^2)$ and at rapidity $y=0$, one has
$x_a=x_b=\sqrt{\tau}=M/\sqrt{s}$, where $M$ is the lepton pair mass.
For the sake of simplicity, by neglecting the sea-sea contributions,
one gets
\begin{equation}\label{ADY2}
A_{DY}=\frac{(4\lambda_v-1)(\lambda_s-1)+(\lambda_v-1)(4\lambda_s-1)}
{(4\lambda_v+1)(\lambda_s+1)+(\lambda_v+1)(4\lambda_s+1)}\ ,
\end{equation}
where $\lambda_v=u_{val}/d_{val}$ and $\lambda_s=\bar u/\bar d$. In a
broad kinematical region where $\lambda_v\sim 2$, this expression shows
the sensitivity of $A_{DY}$ to $\lambda_s$ and, for example, for a
flavor symmetric sea i.e. $\lambda_s=1$, one gets $A_{DY}=0.09$,
whereas $\lambda_s=0.5$ leads to $A_{DY}=-0.09$. This asymmetry has been
measured at CERN by the NA51 Collaboration \cite{Bal} in proton-proton
and proton-deuteron collisions with $450 GeV/c$ incident protons, in
the mass range $4.3<M<8.5 GeV$. As shown in Fig.10, the asymmetry
$A_{DY}$ predicted by our parton densities is compatible with the
result of this rather low statistics experiment. A more accurate
determination of $\lambda_s$ should be obtained in the near future at
FNAL by the E866 experiment \cite{Fermi}. On the other hand, since
there is a realistic possibility of having proton-proton and
proton-deuteron collisions at RHIC with a high luminosity
\cite{RSC,BSNP}, the measurement of $A_{DY}$ should be seriously
envisaged. In view of these future data, we give in Fig.11 the ratio
$(\bar d-\bar u)/$ $(\bar d+\bar u)$ versus $x$ for a standard lepton
pair mass $M=5GeV$ (solid line).

A second possibility for testing the value of $\lambda_s$ has been
proposed in ref.\cite{BSNP} by means of the production of $W^{\pm}$ and
$Z$ in $pp$ and $pn$ collisions, also accessible at RHIC which will
have the requiered energy. For $W^{\pm}$ production, let us consider
the following ratio
\begin{equation}\label{RW}
R_W=\frac{d\sigma^{W^+}_{pp}/dy+ d\sigma^{W^-}_{pp}/dy-
d\sigma^{W^+}_{pn}/dy- d\sigma^{W^-}_{pn}/dy}{d \sigma^{W^+}_{pp}/dy+
d\sigma^{W^-}_{pp}/dy+ d\sigma^{W^+}_{pn}/dy+ d\sigma^{W^-}_{pn}/dy}\ .
\end{equation}
If we neglect the sea-sea contributions, it simply reads in terms of
$\lambda_v(x)$ and $\lambda_s(x)$
\begin{equation}\label{RW2}
R_W=-\frac{(\lambda_v(x_a)-1) (\lambda_s(x_b)-1) + (x_a\leftrightarrow
x_b)}{(\lambda_v(x_a)+1) (\lambda_s(x_b)+1) + (x_a\leftrightarrow
x_b)}\ .
\end{equation}
Clearly $R_W$ is symmetric under $y\to -y$ and $R_W=0$ if the sea is
flavor symmetric, i.e. $\lambda_s(x)=1$. We show in Fig.12 our
predictions for $R_W$ at two different energies. We find that $R_W$ is
positive because $\lambda_s$ is always less than one, as we can check
{}from the dashed line in Fig.11 which corresponds to $Q^2=M^2_W$. $R_W$
decreases when $\sqrt s$ increases because for smaller $x$, according
to Fig.11, $\lambda_s(x)$ increases and it follows from eq.(\ref{RW2})
that $R_W$ gets smaller. Actually the trend shown in Fig.11 is simply
due to the fact that the difference $\bar d-\bar u$ does not change
with $Q^2$, whereas the sum $\bar d+\bar u$ increases with increasing
$Q^2$. The prediction obtained for $R_W$ is slightly different from
that of ref.\cite{BSNP} because we are now using a more reliable set of
parton distributions.

\section{Quark, antiquark, gluon helicity distributions and
experimental tests}
Since our approach is based on the direct construction of the quark
(antiquark) parton distributions of a given helicity $q^{\pm}_{val}$
and $\bar q^{\pm}$, we have nothing to add to the results of section 2
to obtain $\Delta q_{val}=q^+_{val}-q^-_{val}$ and $\Delta \bar q=\bar
q^+-\bar q^-$ at $Q^2=Q^2_0$. We recall the simple relations used in
ref.\cite{BS} namely
\alpheqn
\begin{eqnarray}\label{Delta}
x\Delta u_{val}(x,Q^2_0)&=&xu_{val}(x,Q^2_0)-xd_{val}(x,Q^2_0),\\
x\Delta
d_{val}(x,Q^2_0)&=&-1/3xd_{val}(x,Q^2_0),\label{Deltab}\\
x\Delta
\bar u(x,Q^2_0)&=&x\bar u
(x,Q^2_0)-x\bar d(x,Q^2_0),\label{Deltac}\\
x\Delta
\bar d(x,Q^2_0)&=&x\Delta s
(x,Q^2_0)=x\Delta\bar s(x,Q^2_0)=0\ .\label{Deltad}
\end{eqnarray}
\reseteqn
We show in Fig.13 the different ratios $\Delta q/q$ at $Q^2=3GeV^2$
versus $x$. One finds the usual features for $\Delta u/u$ (positive and
growing at large $x$) and for $\Delta d/d$ (negative and growing at
large $x$), whereas $\Delta\bar u/\bar u$ is very large and negative,
which is less conventional, and $\Delta\bar d/\bar d$ is zero.
Concerning the helicity distribution of the gluon $x\Delta G(x,Q^2_0)$,
in the absence of any serious experimental indication, we can only make
some speculations. One possibility is what one can call the {\it soft
gluon}, for which one assumes
\alpheqn
\begin{equation}\label{DeltaG}
x\Delta G(x,Q^2_0)=x^2G(x,Q^2_0)
\end{equation}
and another one, is the {\it hard gluon}, where one takes
\begin{equation}\label{DeltaGb}
x\Delta G(x,Q^2_0)=\lambda(1-x)x^2G(x,Q^2_0)
\end{equation}
\reseteqn
with $\lambda=4$ which is the maximum value allowed to obey positivity,
i.e. $\Delta G<G$. The integral of $\Delta G(x)$ for the hard gluon
case is about four times larger than for the soft gluon case. Note that
for these two possibilities we assume $\Delta G(x)>0$, which need not
be true. In Fig.14, together with $xG(x)$, we show $x\Delta G(x)$ for
these two choices which do not involve any additional parameter.

Having fixed all these helicity distributions at $Q^2=Q^2_0$, at higher
$Q^2$ they are obtained from the DGLAP evolution equations \cite{GL}.
For the valence contributions, since the splitting functions are the
same for $q_{val}$ and $\Delta q_{val}$, one can evolve indifferently
the r.h.s. or the l.h.s. of eqs.(\ref{Delta},\ref{Deltab}),
 it will lead to the
same result which is, anyway, independent of the gluon helicity
distribution. This is not the case for the antiquarks and in particular
for $\Delta\bar u$, if one evolves the r.h.s. of eq.(\ref{Deltac}), the
contribution coming from $G(x,Q^2)$ will cancel in the difference. This
situation corresponds to the assumption $\Delta G(x,Q^2)\equiv 0$ for
all $Q^2$. On the other hand, if one evolves correctly the l.h.s. of
eq.(\ref{Deltac}), it will depend on $\Delta G(x,Q^2)$ for which we can
make either of the two different choices mentioned above. For the
antiquarks, one can also assume that eq.(\ref{Deltad}) remains valid for
all $Q^2$ or starting from zero at $Q^2=Q^2_0$ one generates a non-zero
$\Delta\bar q$ at higher $Q^2$ from the correct $Q^2$ evolution with
either the soft or the hard gluon. In Fig.15a,b we show the different
ratios $\Delta q/q$ versus $x$ evaluated at $Q^2=M^2_W$ for the three
different situations described above. For $\Delta u/u$ the results are
very similar and $\Delta\bar u/\bar u$ decreases in magnitude for a
harder gluon. This is also the case for $\Delta d/d$, but it is
reversed for $\Delta\bar d/\bar d$. Finally we would like to recall
that for the gluon helicity distribution, due to the QCD evolution, one
obtains \cite{EIN} in the very small $x$ region
\begin{equation}\label{Delta/G}
\Delta G(x,Q^2)/G(x,Q^2)\build{\sim}_{x\to 0}^{}
x\ exp\left[0.8\sqrt{S(Q^2)\ell n 1/x}\right]
\end{equation}
for six quark flavors, where $S(Q^2)=\ell nt/t_0$ and $t=\ell
nQ^2/\Lambda^2$, so at fixed $x$ it grows for large $Q^2$.

We now turn to some of the predictions we can make from these
distributions for helicity dependent observables. Let us start with the
polarized structure functions for proton $g^p_1(x,Q^2)$, for neutron
$g^n_1(x,Q^2)$ and for deuteron $g^d_1(x,Q^2)$. In Fig.16 we have
collected different sets of data for $xg^p_1$, the earlier EMC results
\cite{ASH} and the SMC results \cite{ADA} corresponding to
$<Q^2>=10GeV^2$ and the very recent E143 results \cite{ABE}
corresponding to $<Q^2>=3GeV^2$. We also show for comparison our
theoretical predictions, down to the lowest $x$ range evaluated at
these two $Q^2$ values. The agreement is very satisfactory and gives
strong support to our simple relations
eqs.(\ref{Delta}, \ref{Deltab}, \ref{Deltac}, \ref{Deltad})
 between unpolarized and
polarized parton densities. For the integrals of $g^p_1$ we find
\alpheqn
\begin{equation}\label{Inta}
\int^{0.7}_{0.003}g^p_1(x,Q^2)dx=0.132\quad\hbox{at}\quad Q^2=10GeV^2
\end{equation}
compared to the SMC value \cite{ADA} $0.131\pm 0.011\pm 0.011$ and
\begin{equation}\label{Intb}
\int^{0.8}_{0.029}g^p_1(x,Q^2)dx=0.112\quad\hbox{at}\quad Q^2=3GeV^2
\end{equation}
\reseteqn
compared to the E143 value \cite{ABE} $0.120\pm 0.004\pm 0.008$. In
both cases one finds results, at least two standard deviations, below
the Ellis-Jaffe sum rule \cite{EJ} with QCD corrections \cite{LAR}.

Concerning the neutron polarized structure function $xg^n_1(x)$, we show
in Fig.17 a comparison of the E142 data \cite{ANT} at $Q^2=2GeV^2$ with
our theoretical calculations. The dashed line corresponds to the case
where one would assume that $d$ quarks are not polarized, i.e. $\Delta
d(x)\equiv 0$, and it clearly disagrees with the data. However by
including the $d$ valence quark polarization only according to
eq.(\ref{Deltab}), we obtain the solid line in perfect agreement with the
data and we find for $Q^2=2GeV^2$
\begin{equation}\label{Intg}
\int^{1}_{0}g^n_1(x)dx=-0.020\ .
\end{equation}

Finally for the deuteron polarized structure function $g^d_1$, let us
recall that we have the standard relation
\begin{equation}\label{gd}
g^d_1(x)=\frac{1}{2}\left(g^p_1(x)+g^n_1(x)\right)\left(1-1.5
\omega_D\right)\ ,
\end{equation}
where $\omega_D$ is the $D$-state probability in the deuteron. We show
in Fig.18 our theoretical prediction at $Q^2=3GeV^2$ compared to the
preliminary E143 data \cite{TER} which are more accurate than the
earlier SMC data \cite{ADE}. For the integral in the $x$ range covered
by E143 we find
\begin{equation}\label{Intgd}
\int^{0.8}_{0.029}g^d_1(x)dx=0.043
\end{equation}
in fair agreement with the experimental value \cite{TER} $0.044\pm
0.004\pm 0.004$.

Concerning the important issue of the validity of the Bjorken sum rule
\cite{BJO}, our calculations at $Q^2=5GeV^2$ leads to
\begin{equation}\label{Intgp}
\int^{1}_{0}\left[g^p_1(x,Q^2)-g^n_1(x,Q^2)\right]dx=0.158
\end{equation}
perfectly compatible with the best experimental estimate \cite{ADA}
that is $0.166\pm 0.017$. However this is in slight disagreement with
the present theoretical estimate, including QCD corrections \cite{LV}
up to the third order in $\alpha_s$, which gives $0.185\pm 0.004$. We
recall that, to begin with, we did not impose this constraint on our
distributions.

Before moving to hadronic collisions, we would like to close this
discussion with a few words on the ''transverse'' spin dependent
structure function $g_2(x,Q^2)$ which can be measured in polarized
deep-inelastic scattering with a transversely polarized target. The
properties of $g_2$ have been reviewed recently together with some
estimates \cite{JAF} and we recall that $g_2$ can be thought as a sum
of two terms
\begin{equation}\label{g}
g_2(x,Q^2)=g_2^{WW}(x,Q^2)+\bar g_2(x,Q^2)\ ,
\end{equation}
where the first term is a twist-2 contribution \cite{WW} determined
entirely by $g_1(x,Q^2)$ since we have
\begin{equation}\label{gWW}
g^{WW}_2(x,Q^2)=-g_1(x,Q^2)+\int^{1}_{x}\frac{dy}{y}g_1(y,Q^2)
\end{equation}
The second term in eq.(22) has twist-3 contributions, determined by
quark-gluon interactions, and has a priori, no reason to be small.
Given our present knowledge on $g_1$, we can use eq.(\ref{gWW}) to
evaluate $g^{WW}_2$. This is shown in Fig.19 for the proton case and a
comparison with future experimental results will provide an estimate of
$\bar g_2$.

Next we propose some tests using hadronic collisions, in particular in
the framework of the future spin physics programme at RHIC which is
planned to be used as a polarized $pp$ collider \cite{RSC}. Many
helicity dependent observables have been calculated in ref.\cite{BSNP}
and our purpose here is just to show how to update our earlier
predictions, using this new set of helicity distributions. Let us first
consider the parity-violating helicity asymmetry $A_L$ defined as
\begin{equation}\label{AL}
A_L=\frac{d\sigma_-/dy-d\sigma_+/dy}{d\sigma_-/dy+d\sigma_+/dy}
\end{equation}
In $W^+$ production, it reads
\begin{equation}\label{ALy}
A_L(y)=\frac{\Delta u(x_a,M^2_W)\bar
d(x_b,M^2_W)-(u\leftrightarrow \bar d)}{ u(x_a,M^2_W)\bar
d(x_b,M^2_W)+(u\leftrightarrow \bar d)}\ ,
\end{equation}
assuming the proton $a$ is polarized. For $W^-$ production, the quark
flavors are interchanged. Using our set of polarized quark densities,
we find at $\sqrt{s}=500GeV$, the predictions shown in Fig.20. Near
$y=+1$, $A_L^{W^+}\sim \Delta u/u$ and $A_L^{W^-}\sim \Delta d/d$
evaluated at $x=0.435$, whereas near $y=-1$ $A_L^{W^+}\sim- \Delta \bar
d/\bar d$ and $A_L^{W^-}\sim- \Delta \bar
u/\bar u$ evaluated at $x=0.059$. Therefore the trends in Fig.20 can be
easily compared with the shapes displayed in Figs.15a,b which also show
the sensitivity to the choice of either soft gluon or hard gluon
helicity distribution.

In $pp$ collisions where both proton beams are polarized, there is
another observable which is very sensitive to antiquark polarizations,
that is the parity-conserving double helicity asymmetry $A_{LL}$
defined as
\begin{equation}\label{ALL}
A_{LL}=\frac{d\sigma_{++}/dy+d\sigma_{--}/dy-d\sigma_{+-}/dy-
d\sigma_{-+}/dy}{d\sigma_{++}/dy+d\sigma_{--}/dy+d\sigma_{+-}/dy+
d\sigma/dy}\ .
\end{equation}
In lepton-pair production, it reads
\begin{equation}\label{ALLsum}
A_{LL}=-\frac{\sum_{i}e^2_i\left[\Delta q_i(x_a)\Delta\bar
q_i(x_b)+(x_a\leftrightarrow x_b)\right]}{\sum_{i}e^2_i\left[
q_i(x_a)\bar q_i(x_b)+(x_a\leftrightarrow x_b)\right]}\ ,
\end{equation}
where $e_i$ is the electric charge of the quark $q_i$.

We have calculated $A_{LL}$ at $\sqrt{s}=100GeV$ which seems best
appropriate to the acceptance of the detectors at RHIC and the results,
in the soft gluon case, are shown in Fig.21. We observe that $A_{LL}$
is positive because it is dominated by $\Delta u\Delta\bar u$, where
$\Delta\bar u$ is negative. $A_{LL}$ increases with increasing
lepton-pair $M$ and of course for $\Delta\bar u=\Delta\bar d=0$ we
would have $A_{LL}=0$. Obviously these tests can be extended by many
other examples as shown in ref.\cite{BSNP}, in particular
parity-violating and parity-conserving helicity asymmetries in
$W^{\pm}$, $Z$ production in $pp$ and $pn$ collisions.

\section{Double spin transverse asymmetries $A_{TT}$}
So far we have considered collisions involving only longitudinally
pola\-ri\-zed proton beams, but of course at RHIC, transversely
polarized protons will be available as well \cite{RSC}. This new
possibility is extremely appealing because of recent progress in
understan\-ding transverse spin effects in QCD, both at leading
twist \cite{RAL} and higher twist levels \cite{QIU}. For the case of the
nucleon's helicity, its distribution among the various quarks and
antiquarks can be obtained in polarized deep-inelastic scattering
{}from the measurement of the structure function $g_1(x)$
mentioned above. However this is not possible for the {\it
transversity} distribution $h_1(x)$ which describes the state of
a quark (antiquark) in a transversely polarized nucleon. The
reason is that $h_1(x)$, which measures the correlation between
right-handed and left-handed quarks, decouples from deep-inelastic
scattering. Indeed like $g_1(x),\ h_1(x)$ is leading - twist and it can
be measured in Drell-Yan lepton-pair production with both initial proton
beams transversely polarized \cite{RAL}. Other possibilities have been
suggested \cite{JI} but in the framework of this paper, we will envisage
also  a practical way to determine
$h_1(x)$, by using gauge boson production in $pp$ collisions with
protons transversely polarized. Let us consider the double spin
transverse asymmetry defined as
\begin{equation}\label{ATT}
A_{TT}=\frac{\sigma_{\uparrow\uparrow}-
\sigma_{\uparrow\downarrow}}{\sigma_{\uparrow\uparrow}+
\sigma_{\uparrow\downarrow}}
\end{equation}
where $\sigma_{\uparrow\uparrow}
(\sigma_{\uparrow\downarrow})$ denotes the cross section with
the two initial protons transversely polari\-zed in the same
(opposite) direction. Assuming that the underlying parton
subprocess is quark-antiquark annihilation, we easily find for $Z$
production
\begin{equation}\label{ATTsum}
A_{TT}=\frac{\sum_{i=u,d}\left(b_i^2-a_i^2\right)
\left[ h_1^{q_i}(x_a) h_1^{\bar q_i}(x_b)+(x_a\leftrightarrow
x_b)\right]}{\sum_{i=u,d}\left(a_i^2+b_i^2\right)
\left[ q_i(x_a)\bar q_i(x_b)+(x_a\leftrightarrow
x_b)\right]}\ .
\end{equation}

This result generalizes the case of the lepton-pair production
\cite{RAL} through an off-shell photon $\gamma^{\star}$ corresponding
to $b_i=0$ and $a_i=e_i$ and which gives
\begin{equation}\label{ATTa}
A_{TT}=a_{TT}\frac{\sum_{i}e^2_i\left[h^{q_i}_1(x_a)h^{\bar
q_i}_1(x_b)+(x_a\leftrightarrow x_b)\right]}{\sum_{i}e^2_i\left[
q_i(x_a)\bar q_i(x_b)+(x_a\leftrightarrow x_b)\right]}\ ,
\end{equation}
where $a_{TT}$ is the parton asymmetry which has a simple expression in
the c.m. frame of the lepton-pair \cite{RAL}.

For
$W^{\pm}$ production, which is pure left-handed and therefore
does not allow right-left interference, we expect $A_{TT}=0$,
since in this case $a_i^2=b_i^2$. This result is worth checking
experimentally.

So far there is no experimental data on these
distributions $h_1^q (x)$ (or $h_1^{\overline q} (x)$), but there are
some attempts to calculate them either in the framework of the MIT bag
model \cite{RAL} or by means of QCD sum rules \cite{IK}.
However the  use of positivity yields  to derive
a model-independent constraint on $h_1^q (x)$ which  restricts
substantially the domain of allowed values \cite{SOF}. Indeed one has
obtained
\begin{equation}\label{q}
q (x) + \Delta q (x) \ge 2 | h_1^q (x)
|\ .
\end{equation}
 which is much less trivial than
\begin{equation}\label{q(x)}
q (x) \ge | h_1^q (x) |\ ,
\end{equation}
as proposed earlier in ref. \cite{RAL}.

In the MIT bag model, let us
recall that these distributions read \cite{RAL}
\begin{equation}\label{qf}
q=f^2+g^2,\ \Delta q=f^2-1/3g^2\ \hbox{ and }\
h^q_1=f^2+1/3g^2
\end{equation}
and they  saturate eq.(\ref{q}). In this case, we observe that
$h^q_1(x)\geq\Delta q(x)$ but this situation cannot be very general
because of eq.(\ref{q}). As an example let us assume $h^q_1(x)=2\Delta
q(x)$. Such a relation
cannot hold for all $x$ and we see that eq.(\ref{q}), in particular if
$\Delta q (x) > 0$, implies $q (x) \ge 3 \Delta q (x)$. This is
certainly not satisfied for all $x$ by the present determination of
the $u$ quark helicity distribution, in particular for large $x$ where
$A_1^p (x)$ is large \cite{ASH,ADA,ABE}. The simplifying assumption
$h_1^q (x) =
\Delta q (x)$, based on the non-relativistic quark model, which  has
been used in some recent calculations \cite{BSNP,JI} is also not
acceptable for all $x$ values if $\Delta q(x) < 0$ because of
eq.(\ref{q}). To illustrate the practical use of eq.(\ref{q}), let
us consider eqs.(\ref{Delta}) and (\ref{Deltac}). It is then possible to
obtain the allowed range of values for $h_1^u (x)$, namely
\begin{equation}\label{u}
u(x)-{1 \over 2} d (x) \ge | h_1^u (x) |
\end{equation}
and similarly for $\bar u$.

We now turn to some predictions for $A_{TT}$ which will depend heavily
on our assumptions. Since the $u$ quark term is expected to dominate
because of the charge factor, we assume $h^{\bar d}_1=0$ and we take
the equality sign in eq.(\ref{u}) assuming $h^{u}_1(x)>0$ and
$h^{\bar u}_1(x)<0$. We show in Figs.22a,b our results for
$A_{TT}/a_{TT}$ in dilepton production at two energies $\sqrt{s}=50$
and $100GeV$ for several values of the dilepton mass $M$. The effect is
larger for increasing $M$ because at fixed energy it corresponds to
higher $x$ values where $\Delta u$ and $\Delta\bar u$ are larger, as
well as $h^u_1$ and $h^{\bar u}_1$. Of course, at fixed $M$ the effect
decreases with increasing energy. Finally we show in Fig.23 our
predictions for $A_{TT}$ in $Z$ production at two different energies
$\sqrt{s}=350$ and $500GeV$. Clearly these predictions are only a guide
for future experiments at RHIC which will indeed lead to the actual
determination of $h_1(x)$.

\section{Concluding remarks}
We have presented a new set of quark-parton densities in terms of
Fermi-Dirac distributions depending on {\it nine} free parameters which
were determined from the spin-average structure functions at low $Q^2$.
Some simple relations between unpolarized and polarized densities,
which were postulated earlier, are well supported by recent data on
spin-dependent structure functions. We have also proposed a simple
expression for the gluon density in terms of a Bose-Einstein
distribution with no additional free parameter. We have then used a
straightforward DGLAP $Q^2$ evolution to get access to a broad $x$ and
$Q^2$ range in order to test our approach with a large number of
deep-inelastic
 scattering data. The predictions give a very satisfactory
description of the CCFR, NMC, BCDMS, SLAC data and we also get, in a
natural way, the sharp rise of $F^{ep}_2(x,Q^2)$ for small $x$,
recently observed at HERA by the H$1$ and Zeus Collaborations. So this
is a brillant confirmation of the validity of perturbative QCD, to be
confronted to more accurate future data at HERA. We have also discussed
the relevant question of flavor symmetry breaking of the light sea
quarks (or antiquarks) of the nucleon which has to be measured in a
broader $x$ and $Q^2$ range. To extend the polarized antiquark (or sea
quark) parton densities to large $Q^2$, we have to make a choice for
the polarized gluon distribution and we used two possiblities, the soft
gluon and the hard gluon. None of these is favoured at present and
experiment will have to decide. Predictions were given for future
measurements in hadronic collisions with longitudinally and
transversely polarized proton beams at RHIC which will allow further
tests of our approach in the spin sector, in particular for dilepton
and $W^{\pm},Z$ production.

\section*{Acknowledgments}
We are grateful to G. Altarelli, B. Badelek, J. Feltesse, E. Hughes, J.
Kwicinski, J.F. Laporte, R. Roberts, W. Stirling, Y. Terrien, M.
Virchaux and R. Windmolders for providing useful informations and for
helpful discussions.

\section*{Figure Captions}

\begin{goodlist}{Fig.22b}

\item[Fig.1a] The structure function $x F_3^{\nu N} (x, Q^2)$ versus
$x$. Data are from ref.\cite{Q} at $Q^2 = 3 GeV^2$, the solid line is
the result of our fit and the dashed curve represents the MRS (A)
parametrization at $Q^2 = 4 GeV^2$ from ref.\cite{MRS}.

\item[Fig.1b] The antiquark contribution $x \bar q (x) = x \bar u (x) +
x \bar d (x) + x \bar s (x)$ at $Q^2 = 3 GeV^2$ (full circles) and
$Q^2 = 5 GeV^2$ (full triangles) versus $x$. Data are from
ref.\cite{F}, the solid line is the result of our fit and the dashed
curve represents the MRS (A) parametrization at $Q^2 = 4 GeV^2$ from
ref.\cite{MRS}.

\item[Fig.2a] The difference $F_2^p (x) - F_2^n (x)$ at $Q^2 = 4 GeV^2$
versus $x$. Data are from ref.\cite{AR}, the solid line is the result
of our calculation and the dashed curve represents the MRS (A)
parametrization at $Q^2 = 4 GeV^2$ from ref.\cite{MRS}.

\item[Fig.2b] The ratio $F_2^n (x) / F_2^p (x)$ at $Q^2 = 4 GeV^2$
versus $x$. Data are from ref.\cite{AR}, the solid line is the result
of our calculation and the dashed curve represents the MRS (A)
parametrization at $Q^2 = 4 GeV^2$ from ref.\cite{MRS}.

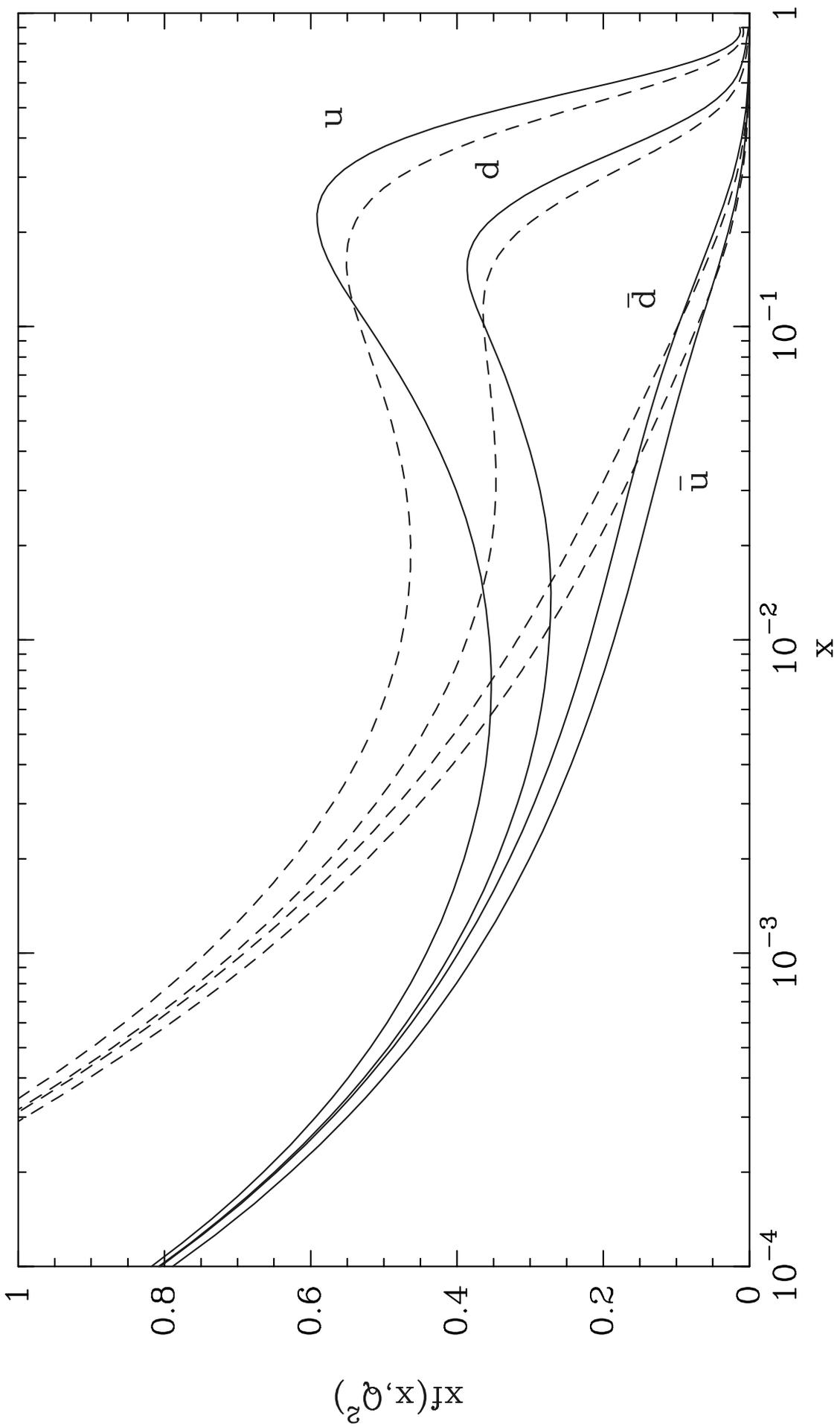
\item[Fig.3] The unpolarized parton distributions as a function of $x$
at two different $Q^2$ values, $Q^2 = 3 GeV^2$ (solid lines) and $Q^2 =
20 GeV^2$ (dashed lines).

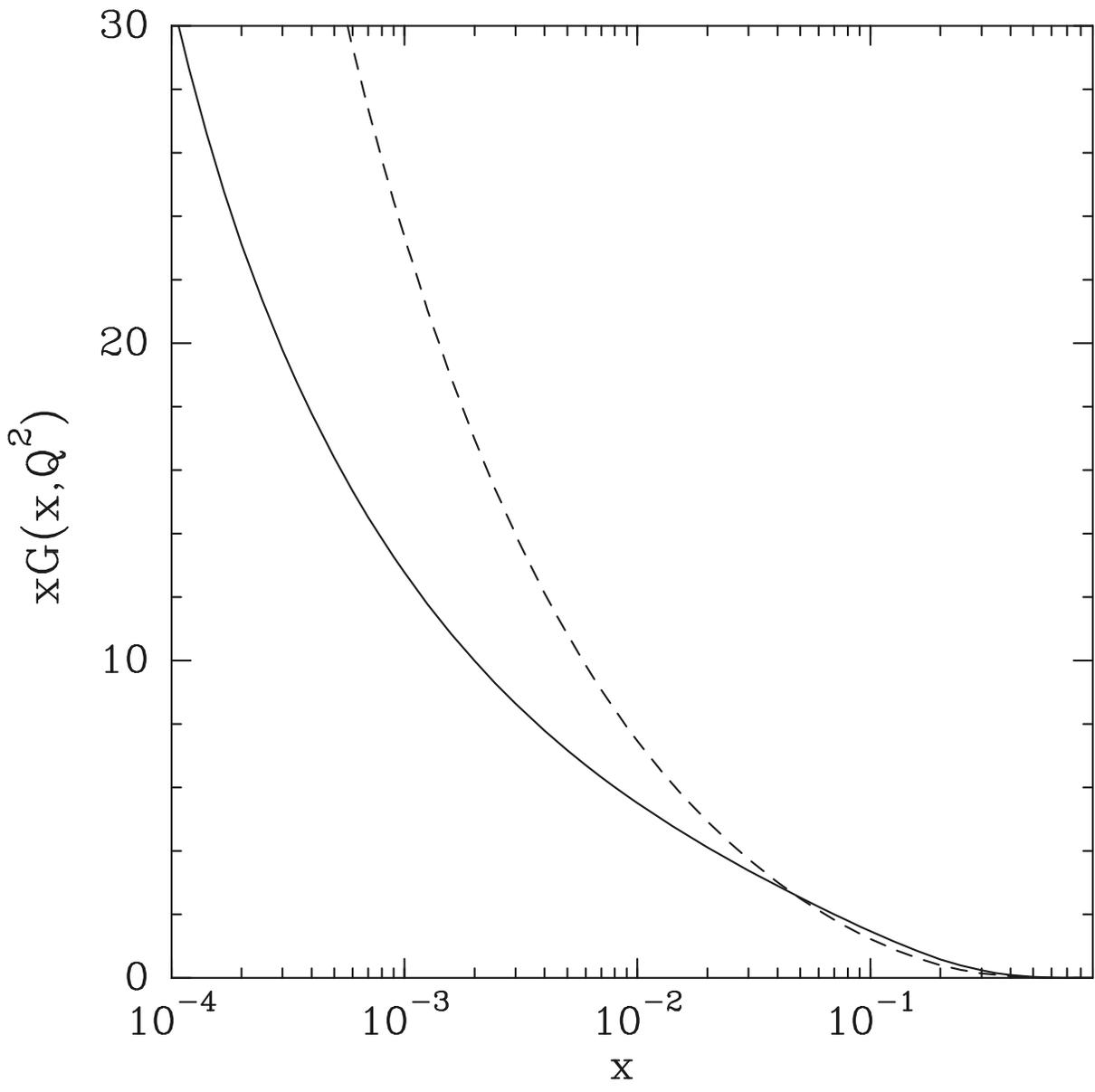
\item[Fig.4] The gluon density $x G (x, Q^2)$ versus $x$ at two
different $Q^2$ values, $Q^2 = 3 GeV^2$ (solid line) and $Q^2 =
20 GeV^2$ (dashed line).

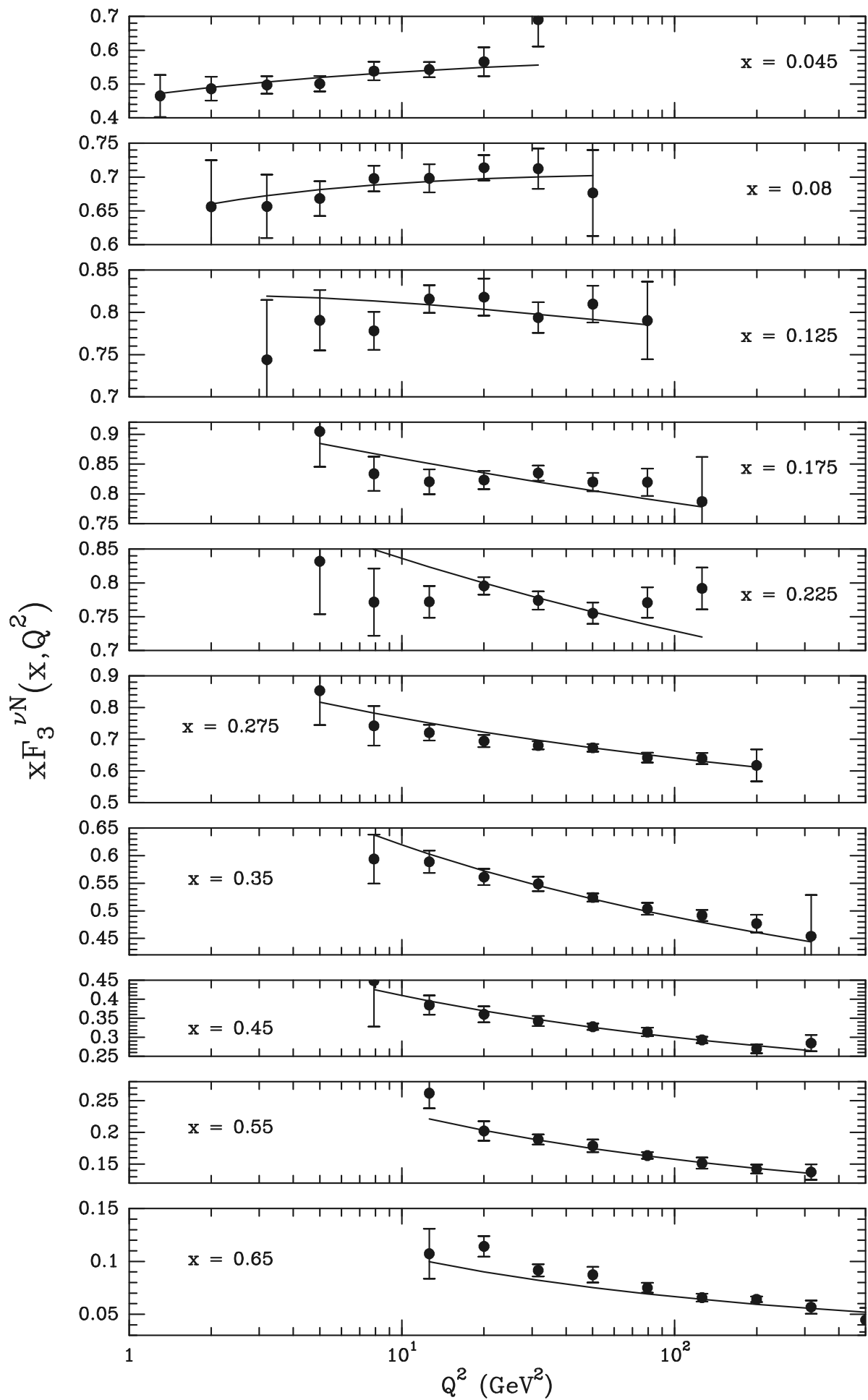
\item[Fig.5] The description of the CCFR data~\cite{Q} of the $x
F_3^{\nu N} (x, Q^2)$ structure function, given by our parton densities.
The theoretical curves are shown after an overall renormalization by a
factor $1.04$.

\item[Fig.6a] The description of the NMC data~\cite{AM} (full circles)
and BCDMS data~\cite{B} (open circles) of the $F_2^{\mu
p} (x, Q^2)$ structure function given by our parton densities versus
$Q^2$ for $x$ bins between $x = 0.0125$ and $x = 0.14$. The theoretical
curves are shown after an overall renormalization by a factor $1.12$.

\item[Fig.6b] Same as Fig.6a for $x$ bins between $x = 0.18$ and $x =
0.75$.

\item[Fig.7a] The description of the Zeus data~\cite{D} (full triangles)
and H$1$ data~\cite{BR} (full circles) of the $F_2^{ep} (x, Q^2)$
structure function given by our parton densities versus $x$ for $Q^2$
bins between $Q^2 = 4.5 GeV^2$ and $Q^2 = 25 GeV^2$.

\item[Fig.7b] Same as Fig.7a for $Q^2$ bins between $Q^2 = 35 GeV^2$ and
$200 GeV^2$.

\item[Fig.7c] Same as Fig.7a for $Q^2$ bins between $Q^2 = 250 GeV^2$
and $5000 GeV^2$.

\item[Fig.8a] The description of the NMC data~\cite{AM} (full circles)
and SLAC data \cite{W} (open circles) of the $F_2^{\mu
p} (x, Q^2)$ structure function in the low $Q^2$, low $x$ region given
by our parton densities for $x$ bins between $x = 0.00375$
and $x = 0.0175$. The theoretical curves are shown after an overall
renormalization by a factor $1.12$.

\item[Fig.8b] Same as Fig.8a for $x$ bins between $x = 0.025$ and $x =
0.14$.

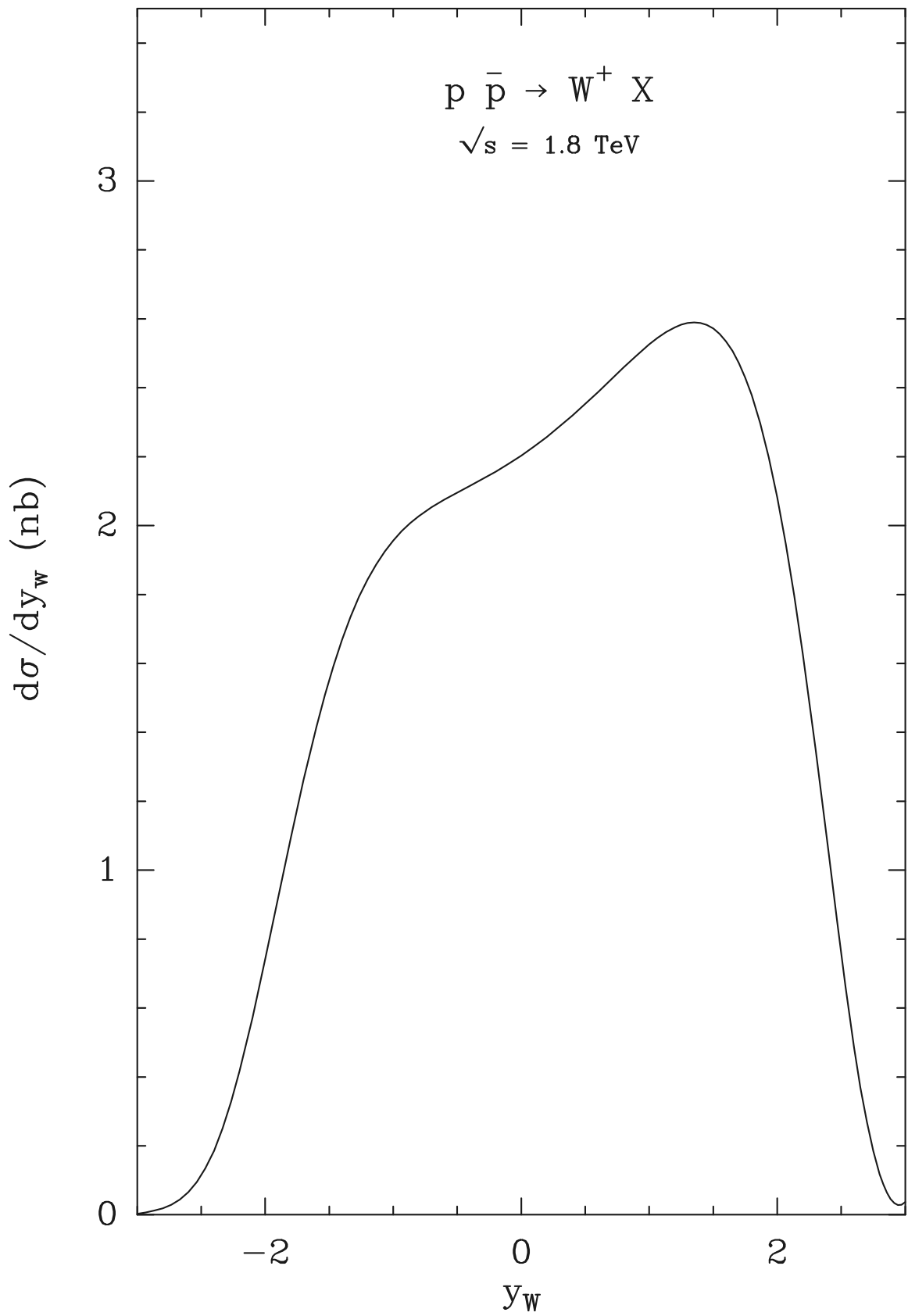
\item[Fig.9] The $W^+$ rapidity distribution in $\bar p p$ collisions
at $\sqrt s = 1.8 TeV$ predicted by our parton densities.

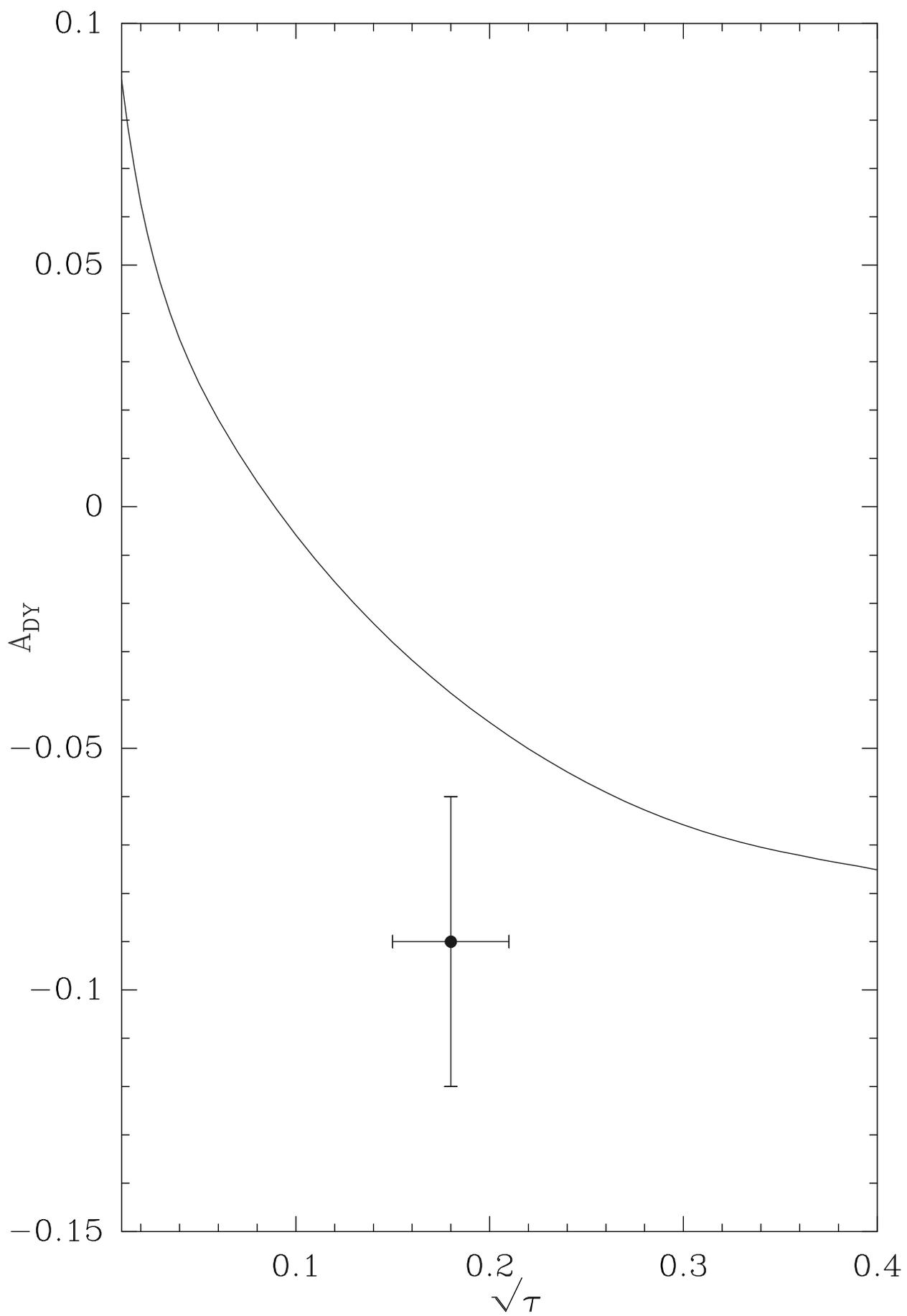
\item[Fig.10] Our theoretical prediction for $A_{DY}$ (see
eq.(\ref{ADY})) versus $\sqrt{\tau} = M / \sqrt s$ compared to the NA51
data~\cite{Bal}.

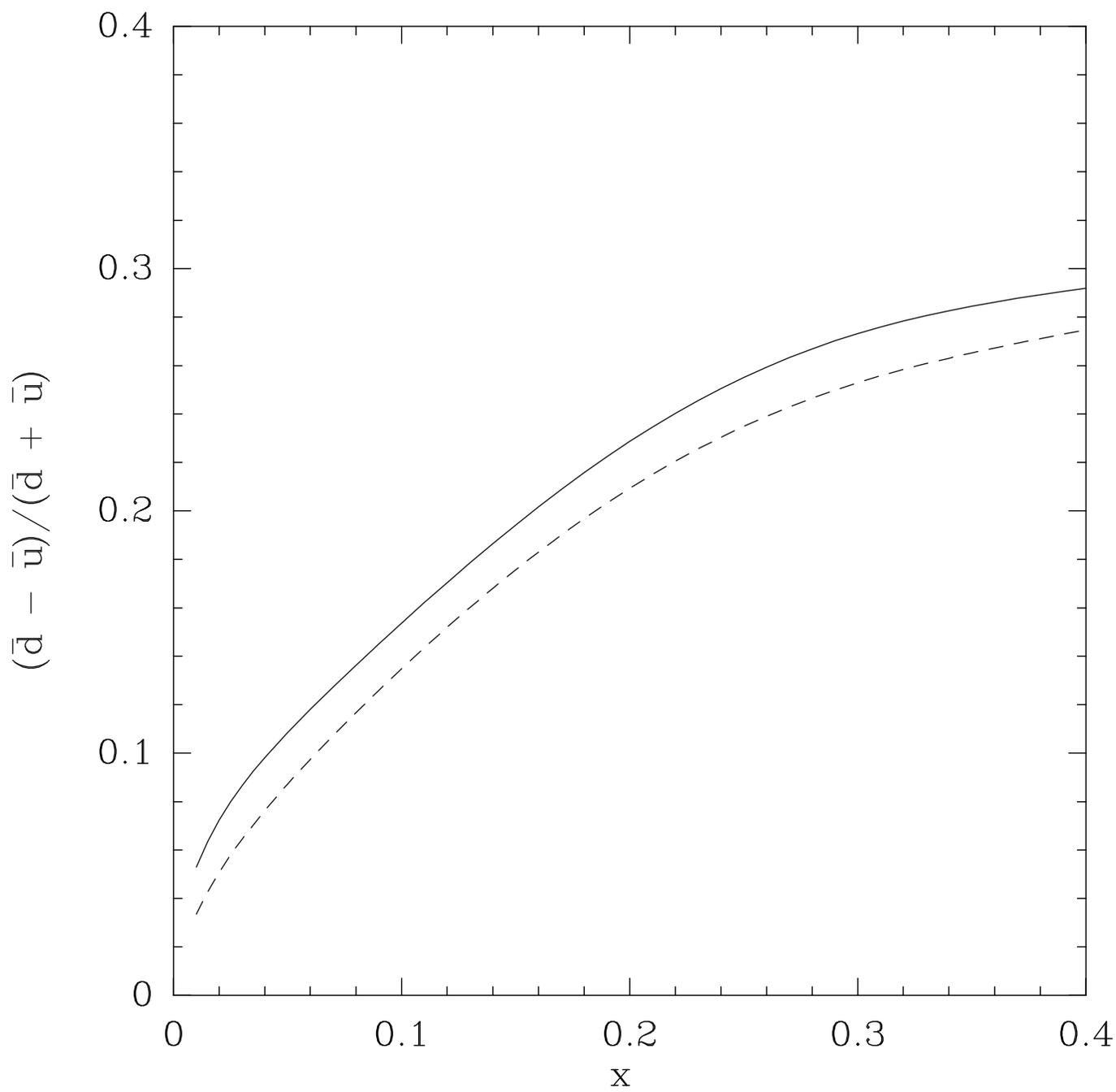
\item[Fig.11] Our predicted ratio $(\bar d - \bar u) / (\bar d + \bar
u)$ versus $x$ for two different $Q^2$ values~; solid line $Q^2 = 25
GeV^2$ and dashed line $Q^2 = M_W^2$.

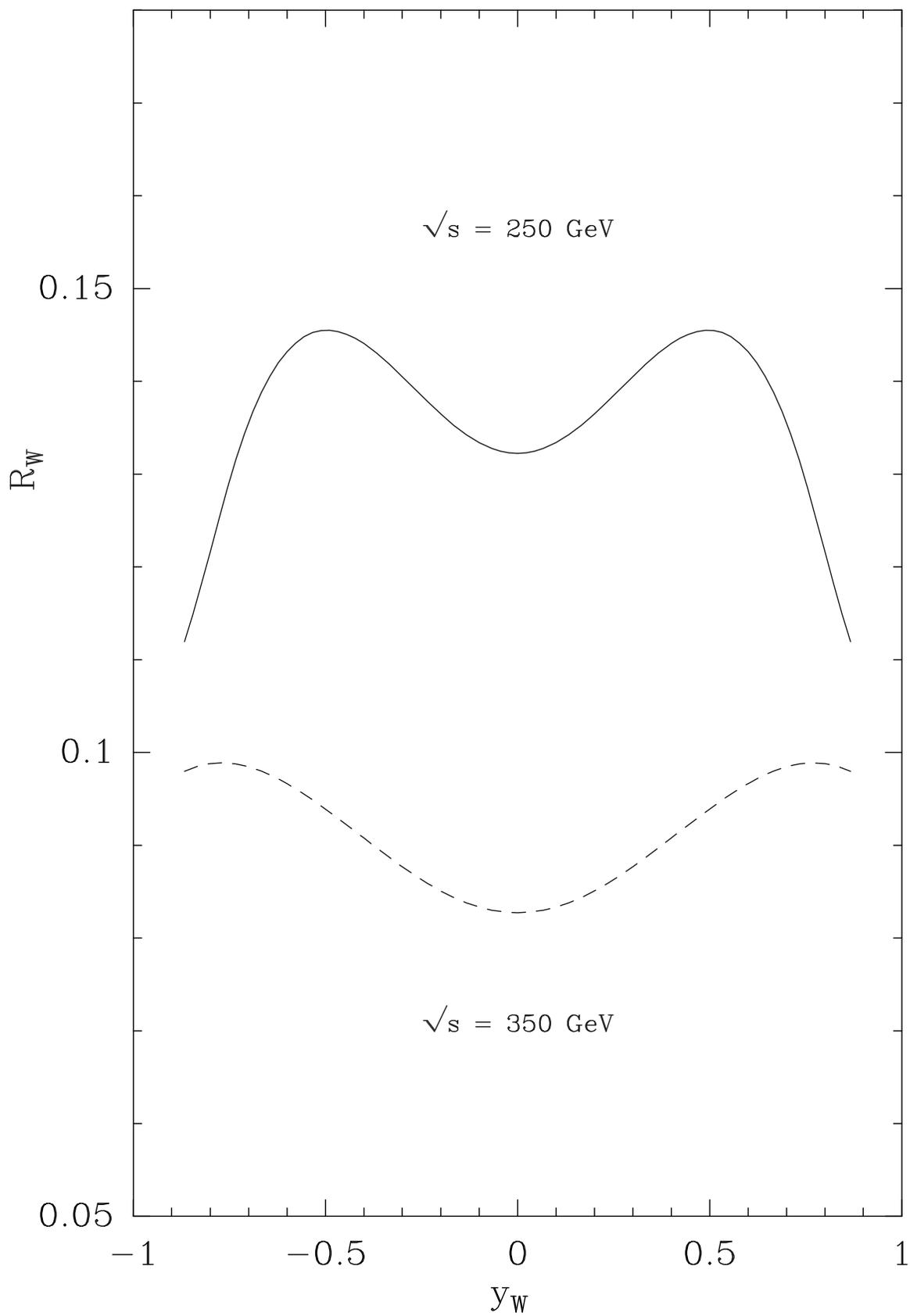
\item[Fig.12] The ratio $R_W$ (see eq.(\ref{RW}))
versus $y$ at $\sqrt s = 250$ and $350 GeV$.

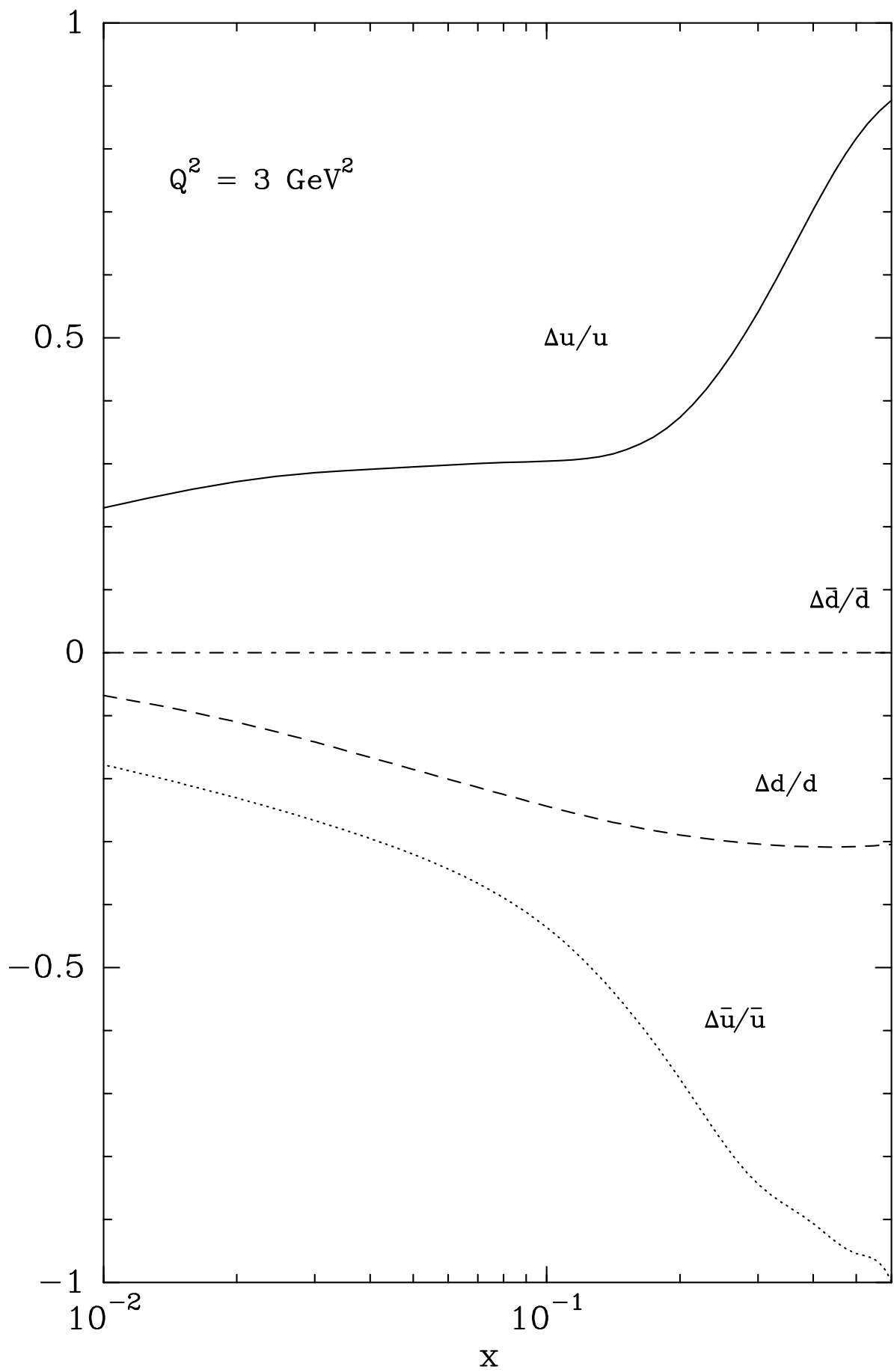
\item[Fig.13] The ratios $\Delta u / u$, $\Delta d / d$, $\Delta \bar u /
\bar u$ and $\Delta \bar d / \bar d$ versus $x$ at $Q^2 = 3 GeV^2$.

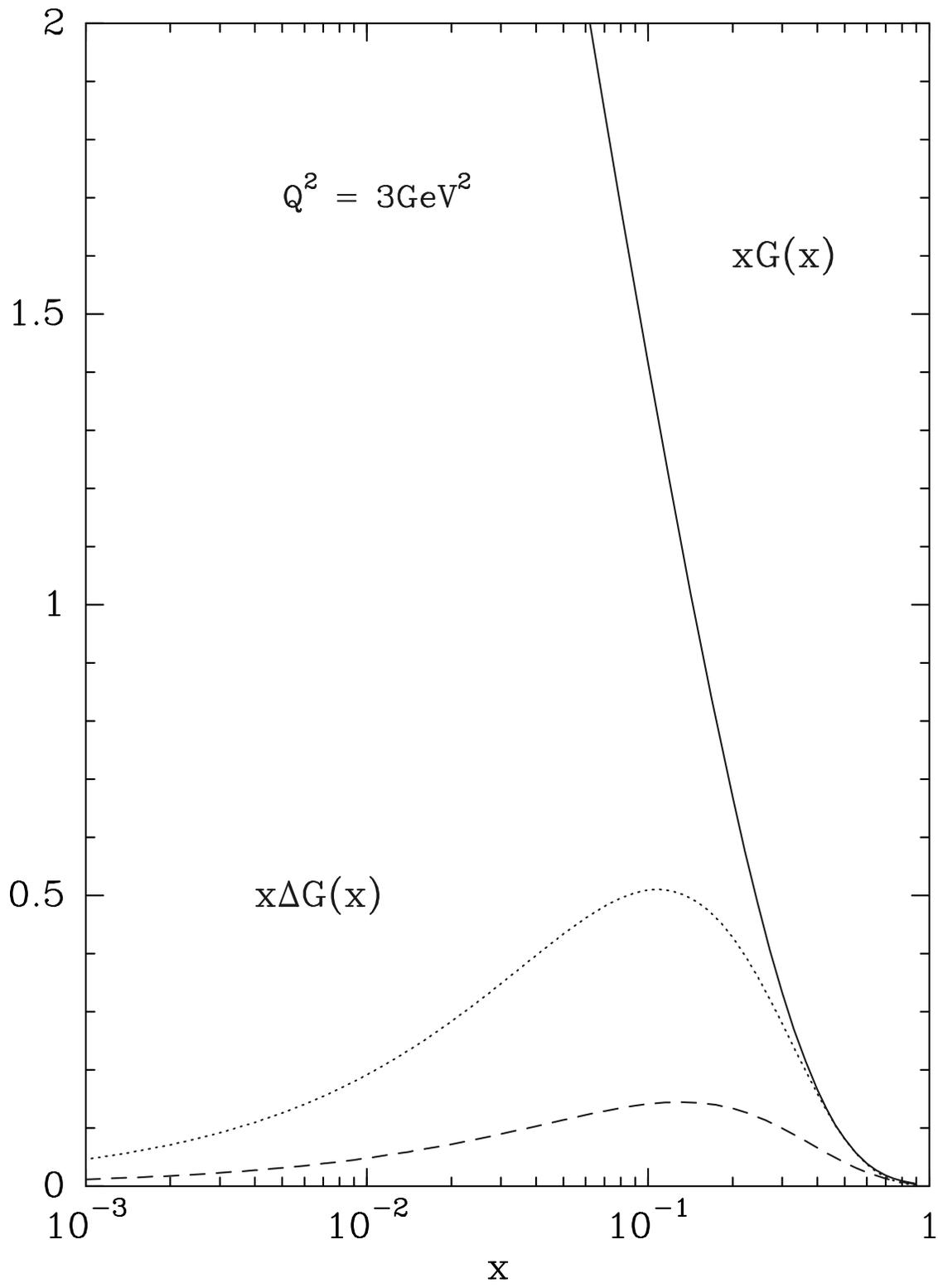
\item[Fig.14] The helicity gluon distribution $x \Delta G (x)$ (soft
gluon, dashed line and hard gluon, dotted line) versus $x$ at $Q^2 = 3
GeV^2$ and for comparison the unpolarized gluon distribution $x G (x)$
(solid line).

\item[Fig.15a] The predicted ratios $\Delta u / u$ and $\Delta \bar u /
\bar u$ versus $x$ at $Q^2 = M_W^2$. The three curves correspond to the
three cases for $\Delta G$: identically zero (solid line), soft gluon
(eq.(\ref{DeltaG})) (dashed-dotted line) and hard gluon
(eq.(\ref{DeltaGb})) (dashed line).

\item[Fig.15b] Same as Fig.15a for $\Delta d / d$ and $\Delta \bar d /
\bar d$.

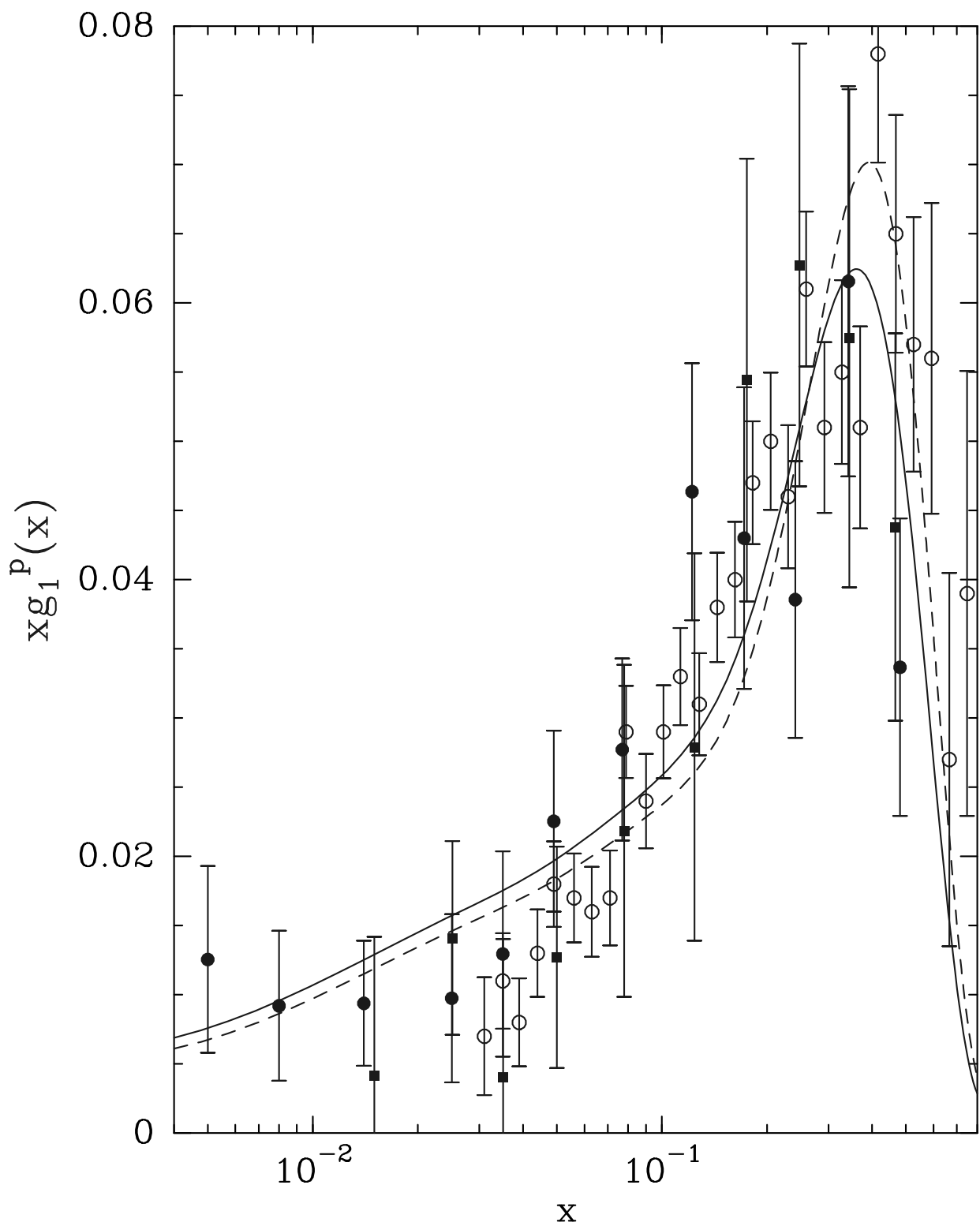
\item[Fig.16] $x g_1^p (x)$ versus $x$ at $< Q^2 > = 3 GeV^2$ and
$10 GeV^2$. Data are from ref.\cite{ASH} (full squares), ref.\cite{ADA}
(full circles) and ref.\cite{ABE} (open circles). The curves are
theoretical predictions at $Q^2 = 3 GeV^2$ (dashed line) and $Q^2 = 10
GeV^2$ (solid line).

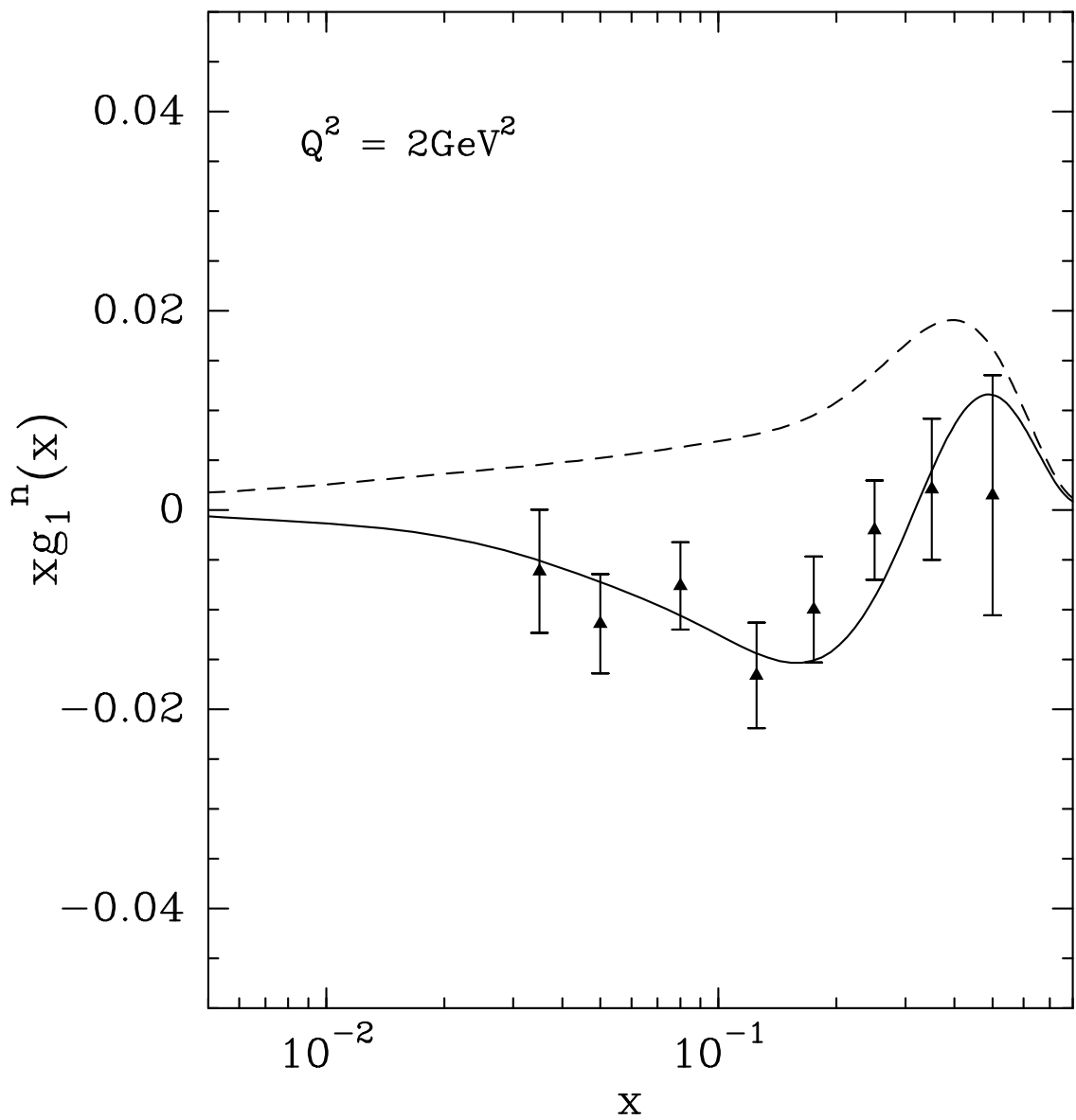
\item[Fig.17] $x g_1^n (x)$ versus $x$ at $< Q^2 > = 2 GeV^2$. Data are
{}from ref.\cite{ANT} together with our theoretical predictions (dashed
line is the contribution of $\Delta u (x)$ and $\Delta \bar u (x)$ only
and solid line is the contribution of $\Delta u (x)$, $\Delta \bar u
(x)$ and $\Delta d_{val} (x)$).

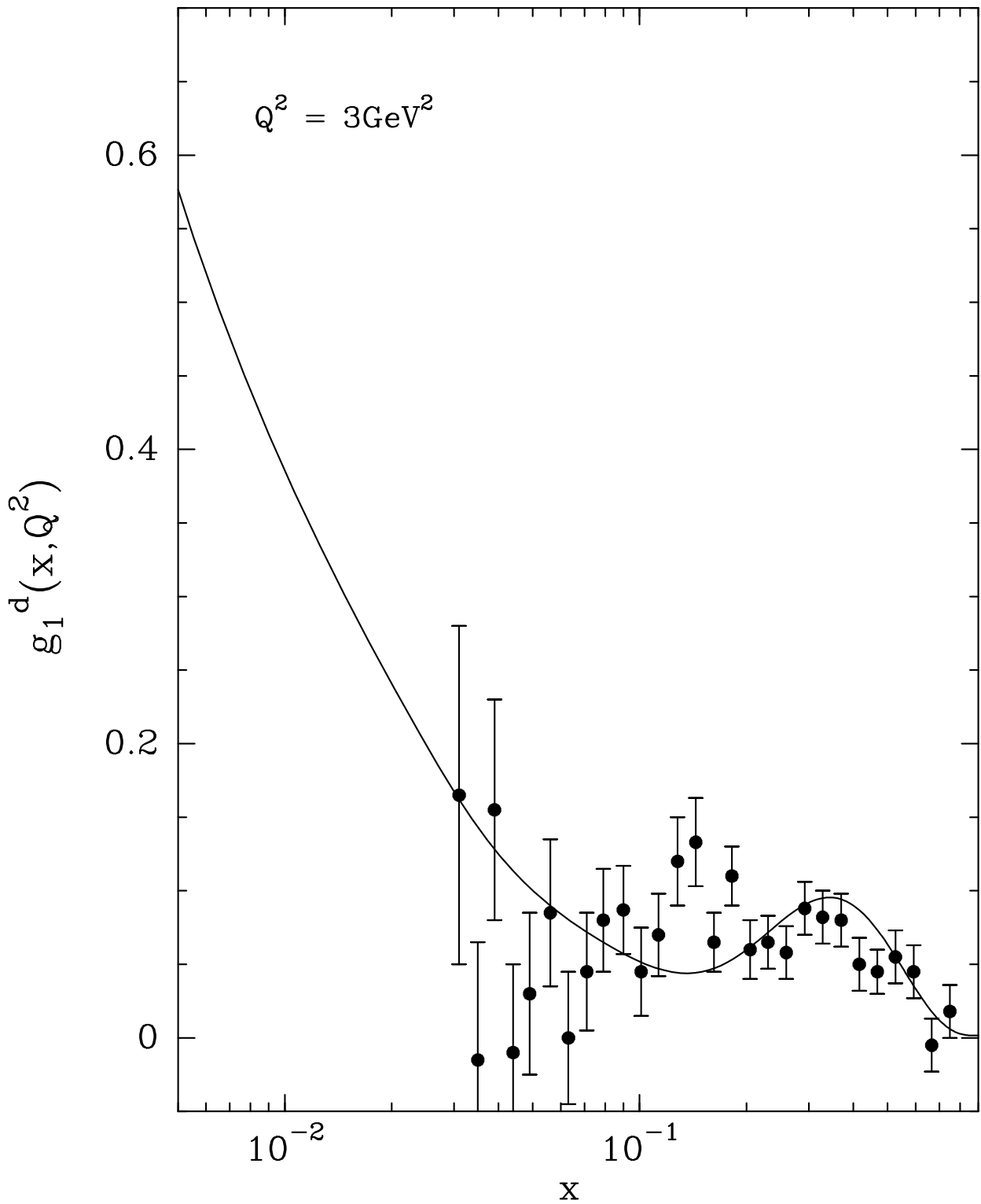
\item[Fig.18] $x g_1^d (x)$ versus $x$ at $< Q^2 > = 3 GeV^2$.
Prelimining data are from ref.\cite{TER} together with our theoretical
prediction.

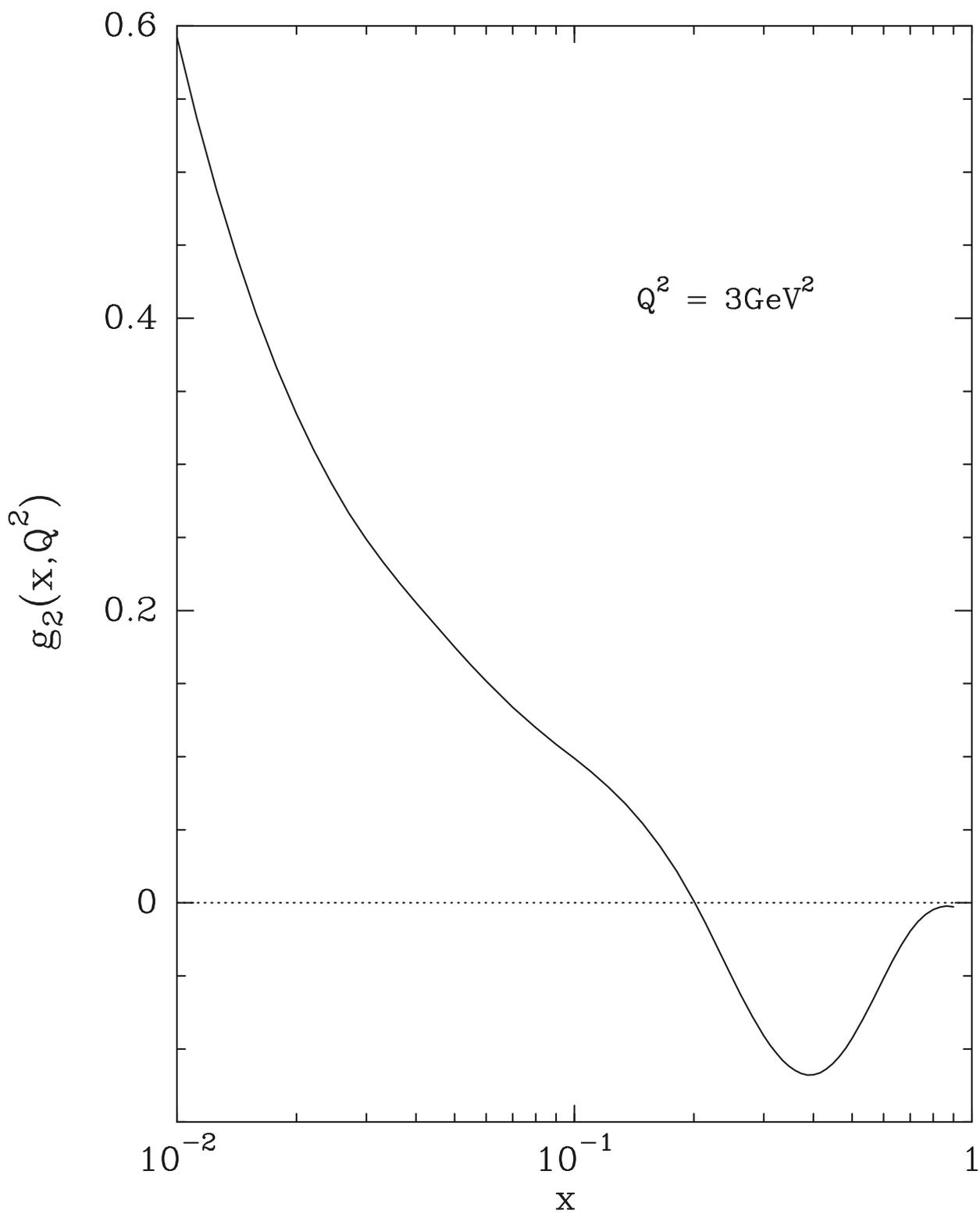
\item[Fig.19] Our theoretical prediction for $g_2 (x, Q^2)$ at $Q^2 = 3
GeV^2$ for proton target assuming $\bar g_2 = 0$ (see eq.(\ref{g})).

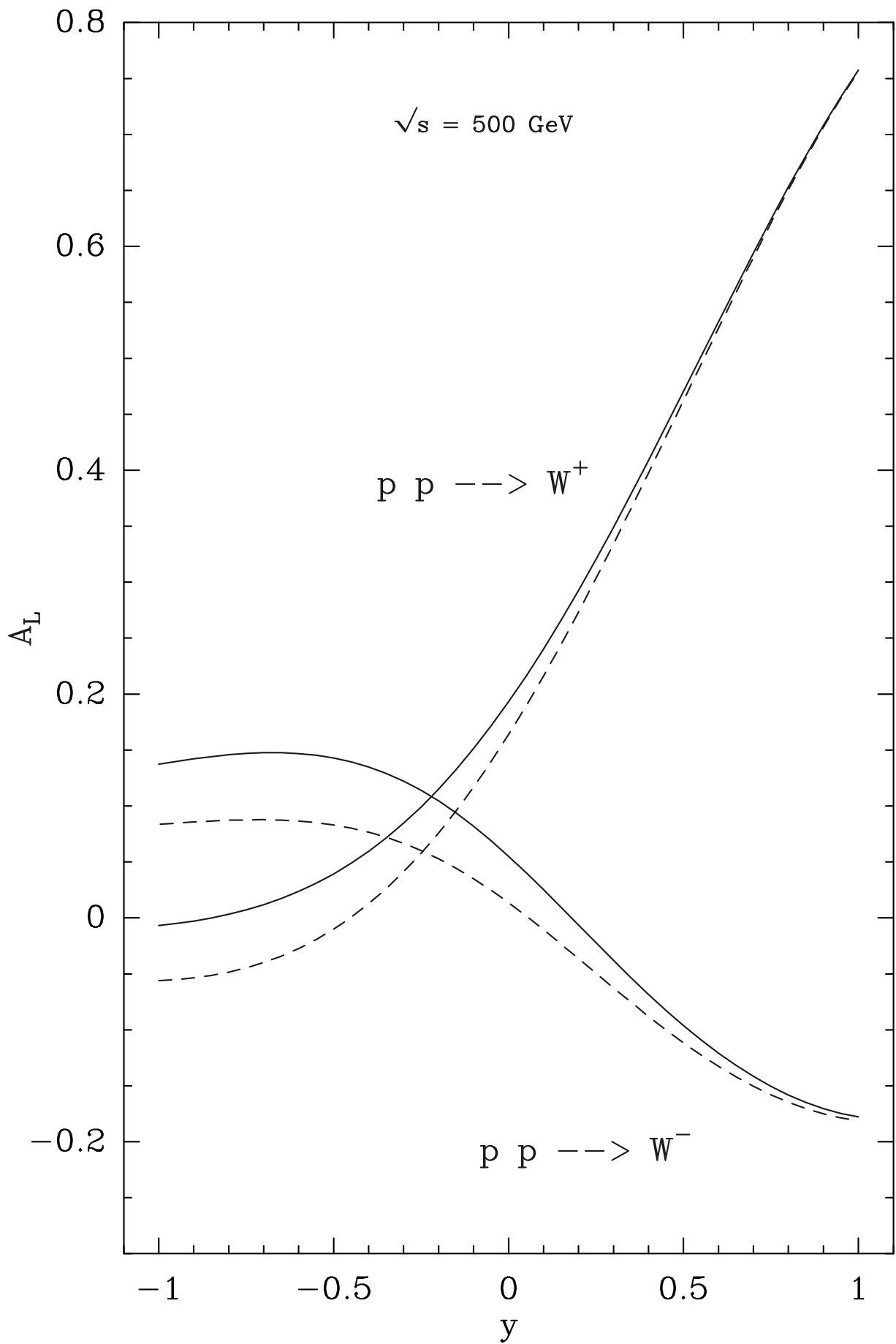
\item[Fig.20] Parity-violating helicity asymmetry $A_L$ versus $y$ for
$W^{\pm}$ production in $pp$ collisions at $\sqrt s = 500 GeV$. Solid
lines correspond to soft gluon and dashed lines to hard gluon.

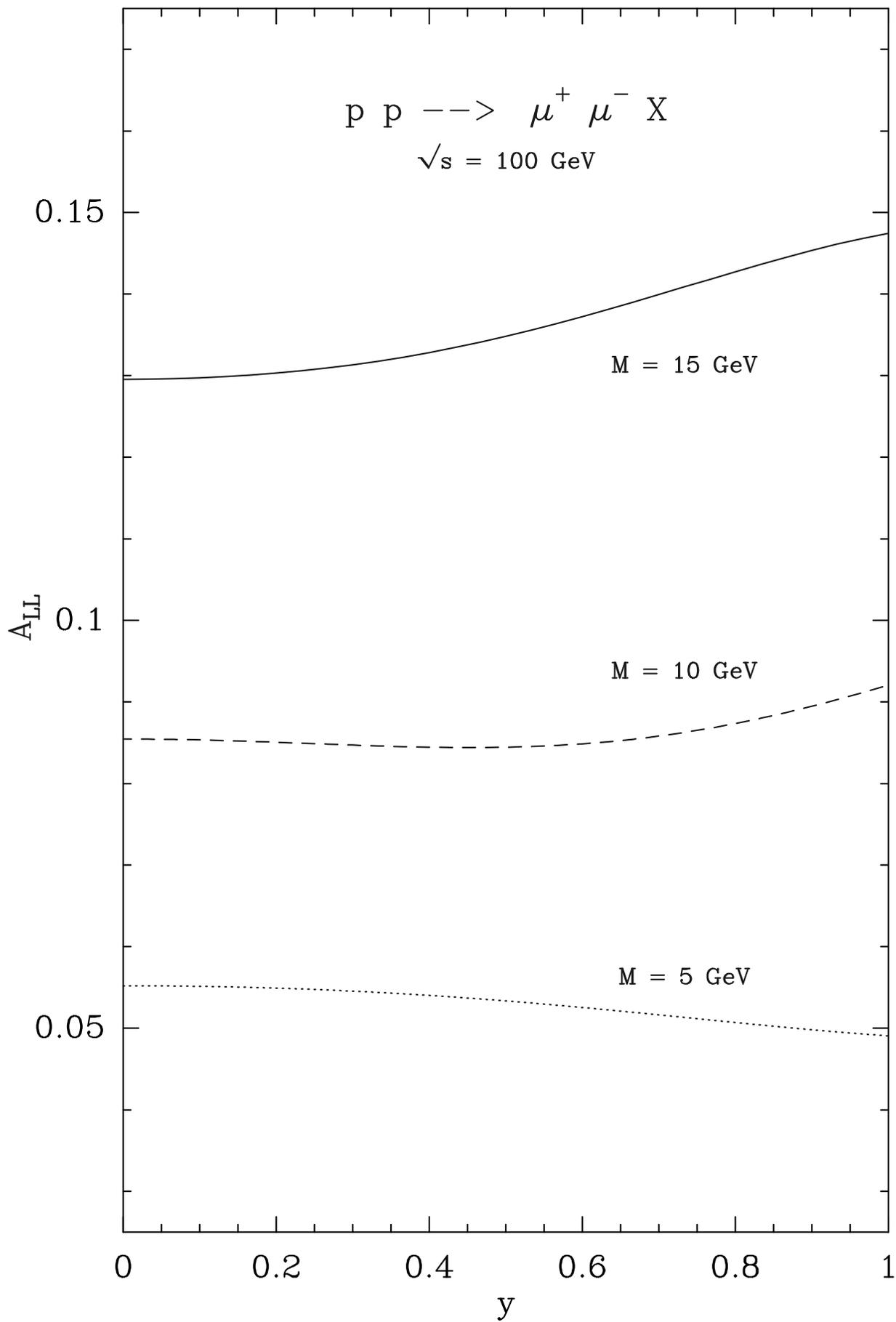
\item[Fig.21] Parity-conserving helicity asymmetry $A_{LL}$ versus $y$
for dilepton production at $\sqrt s = 100 GeV$ and different values of
the lepton-pair mass. The curves correspond to the soft gluon case.

\item[Fig.22a] Predictions for $A_{TT} / a_{TT}$ in dilepton production
versus $y$ for $\sqrt s = 50 GeV$ and $M = 5$, $10$, $15 GeV$. The
curves correspond to the soft gluon case.

\item[Fig.22b] Same as Fig.22a for $\sqrt s = 100 GeV$.

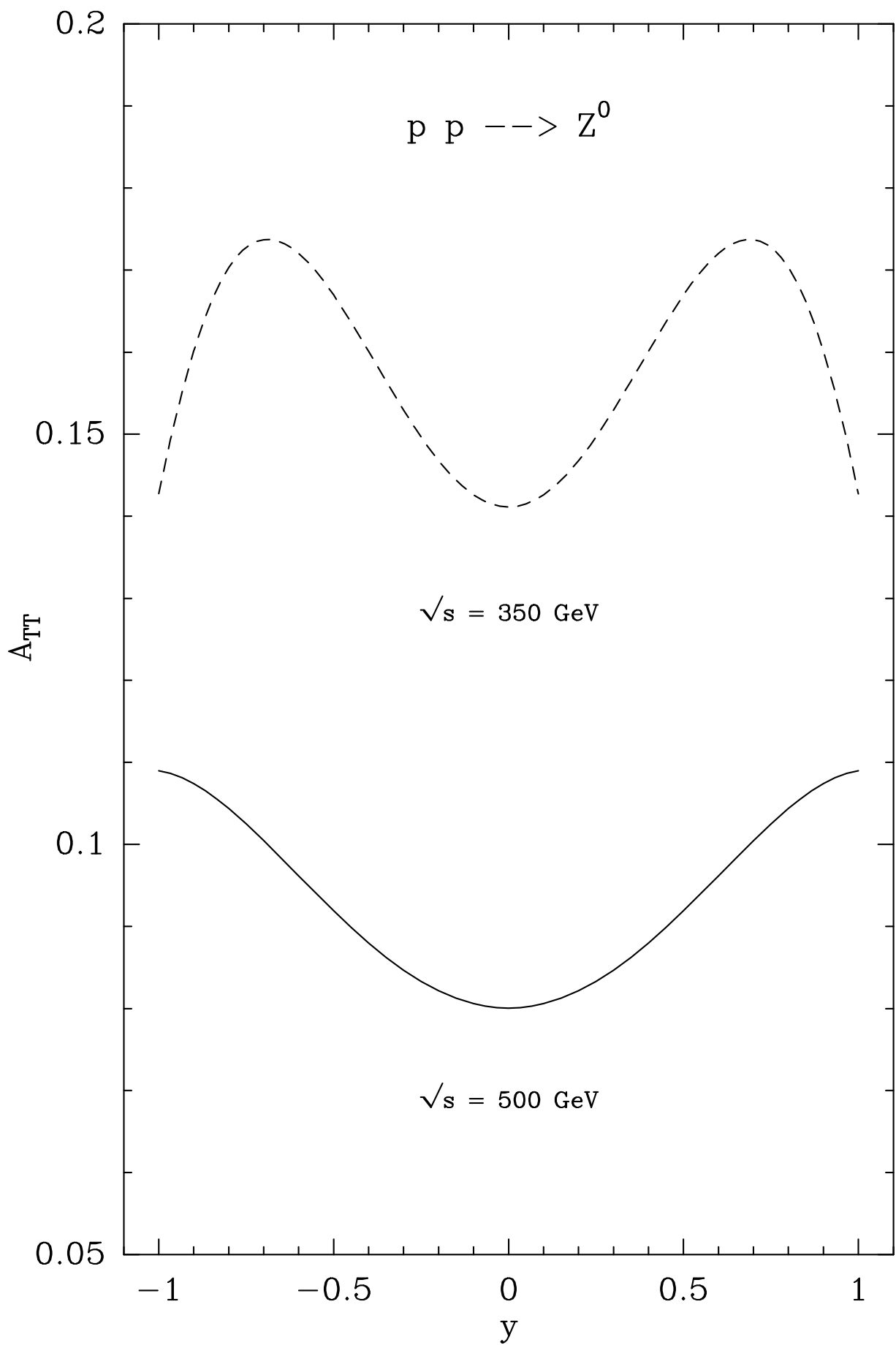
\item[Fig.23] Predictions for $A_{TT}$ in $Z$ production
versus $y$ for $\sqrt s = 350$ and $500 GeV$. The curves correspond to
the soft gluon case.

\end{goodlist}

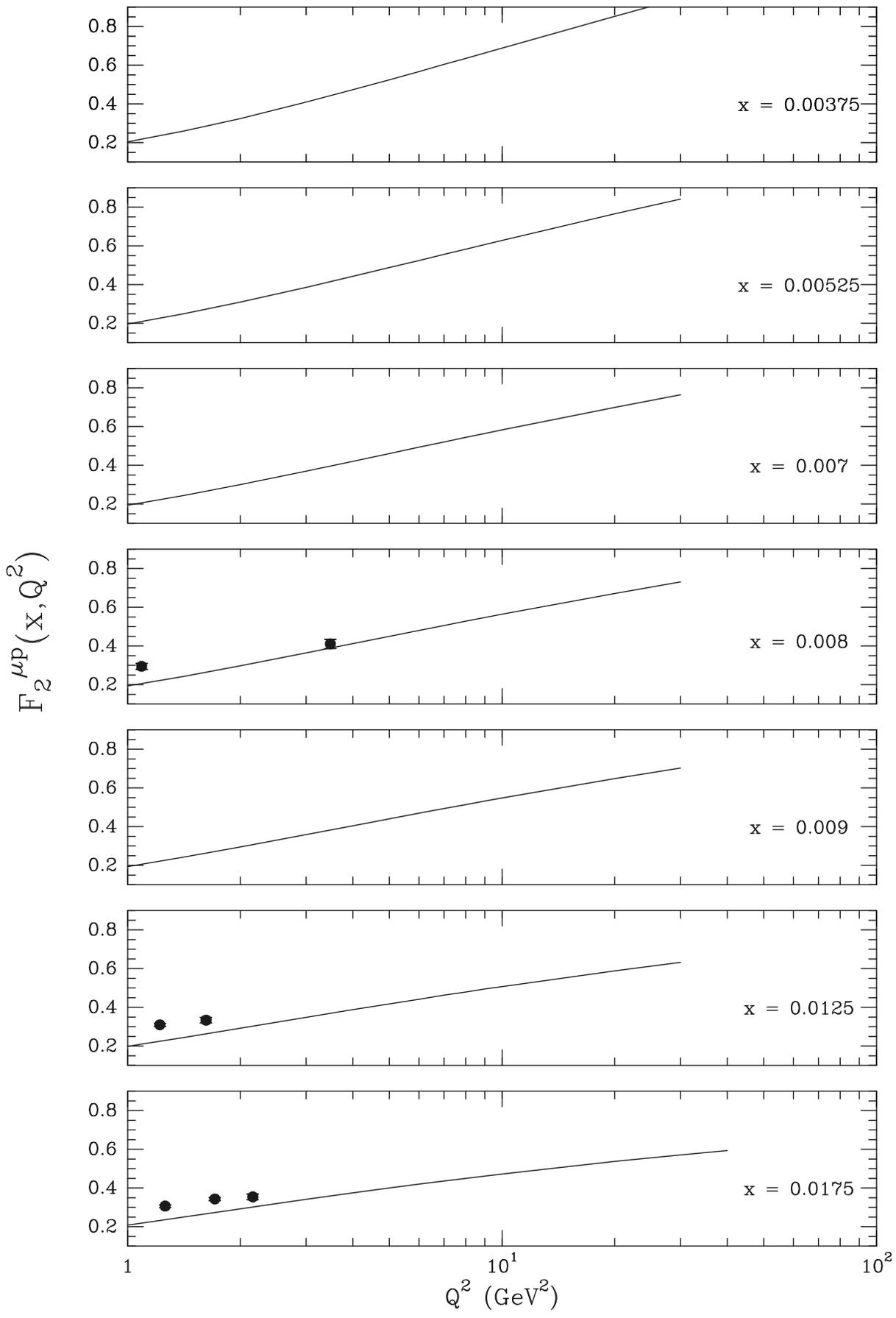
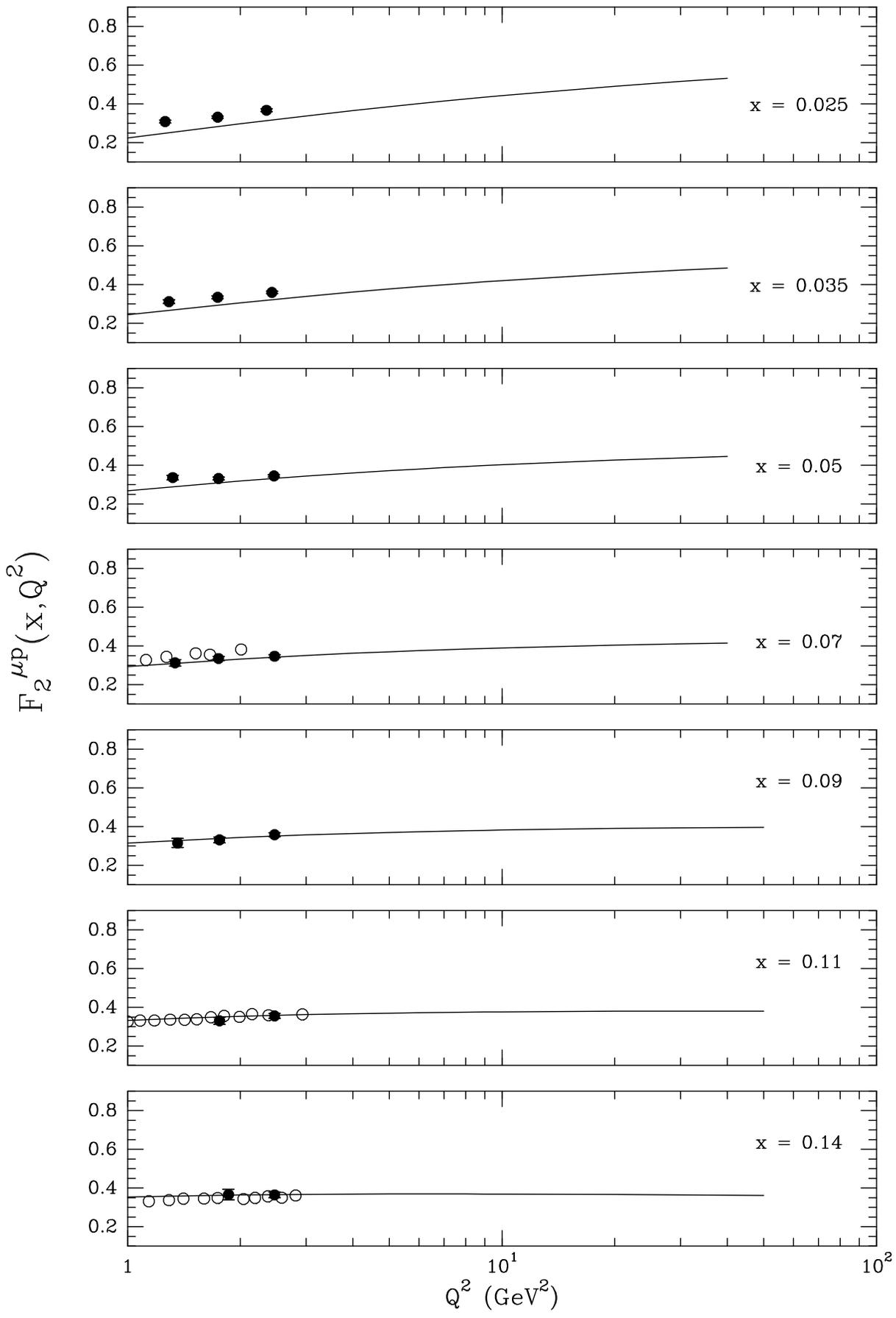
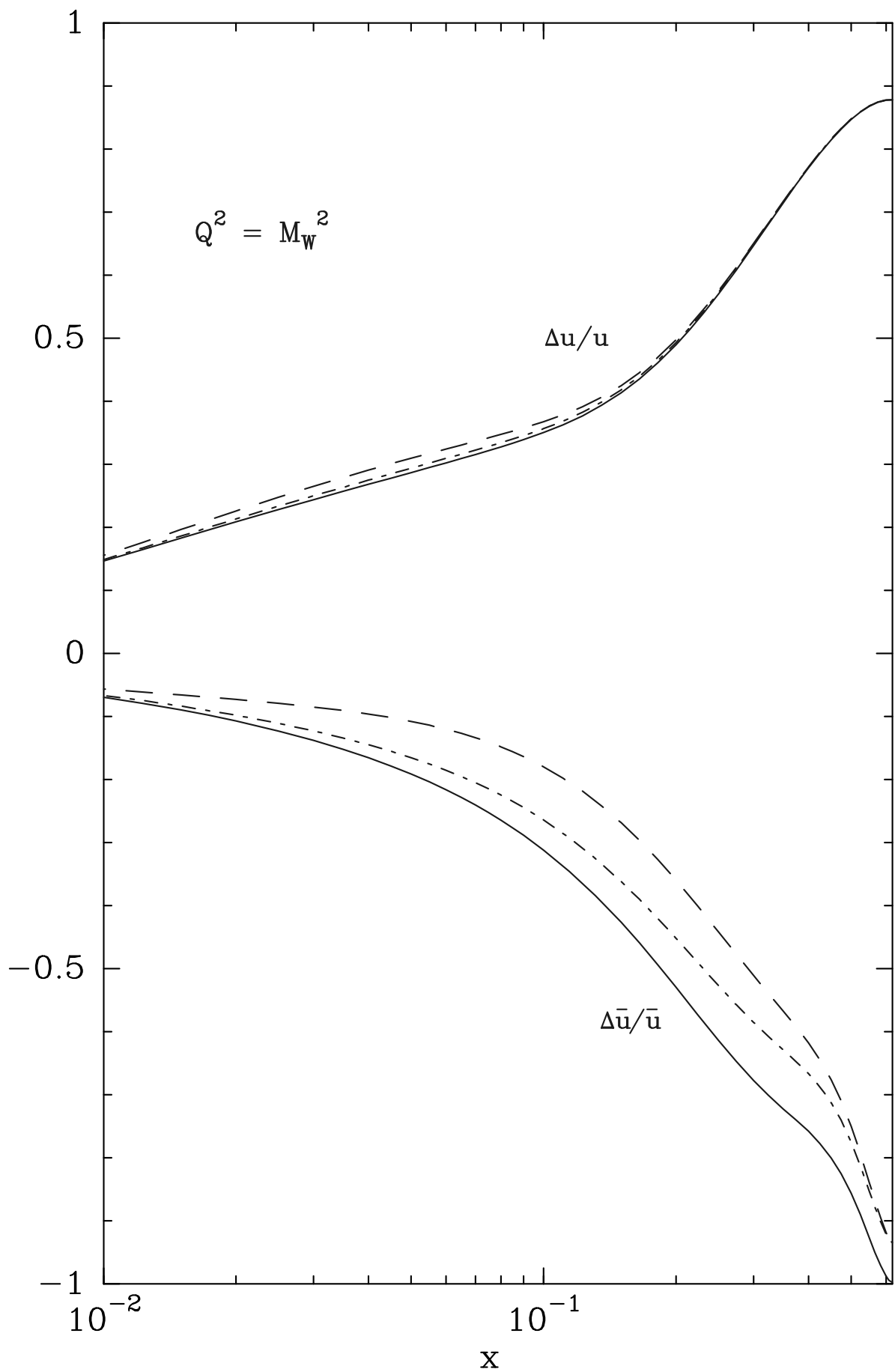
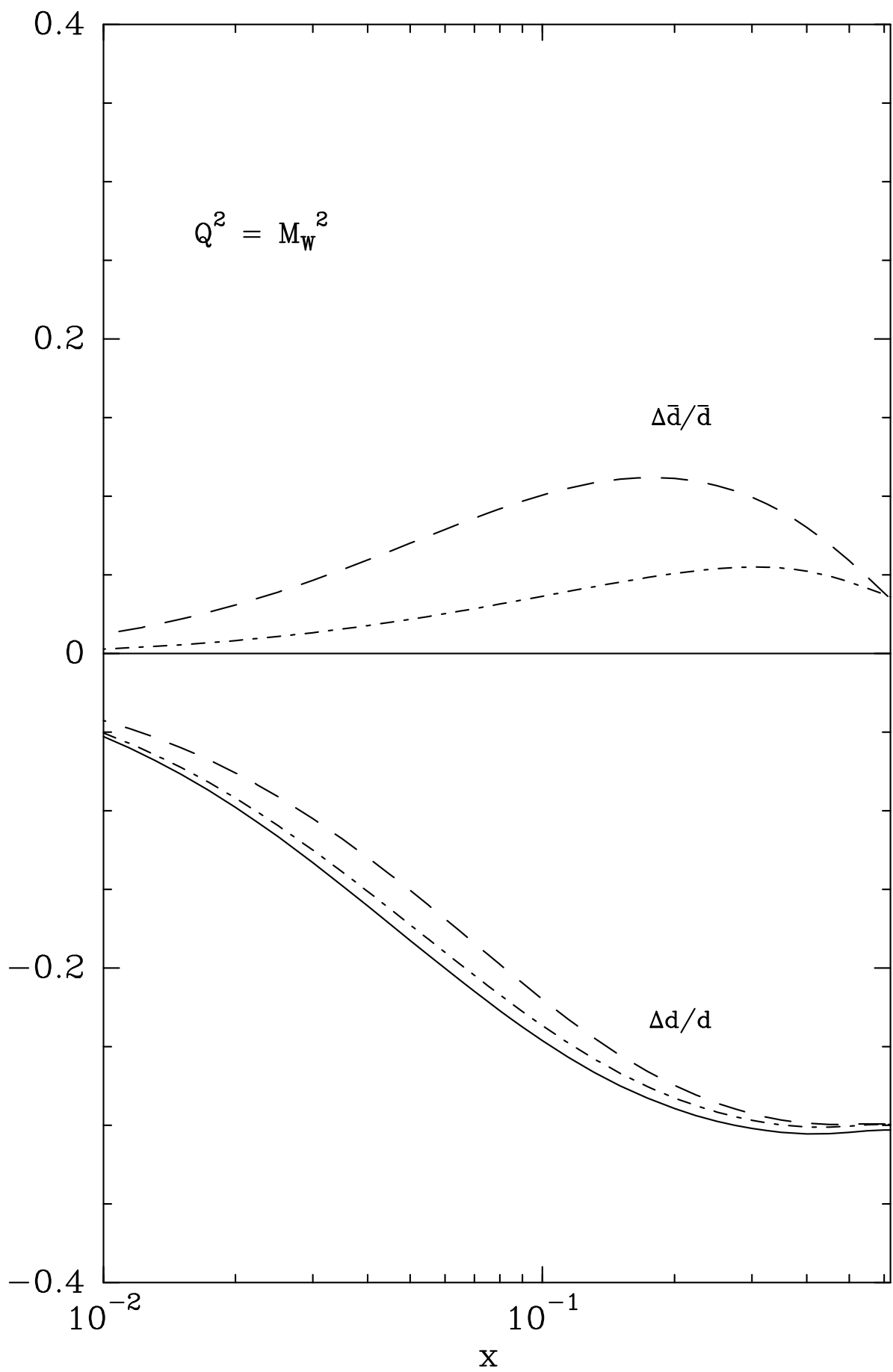
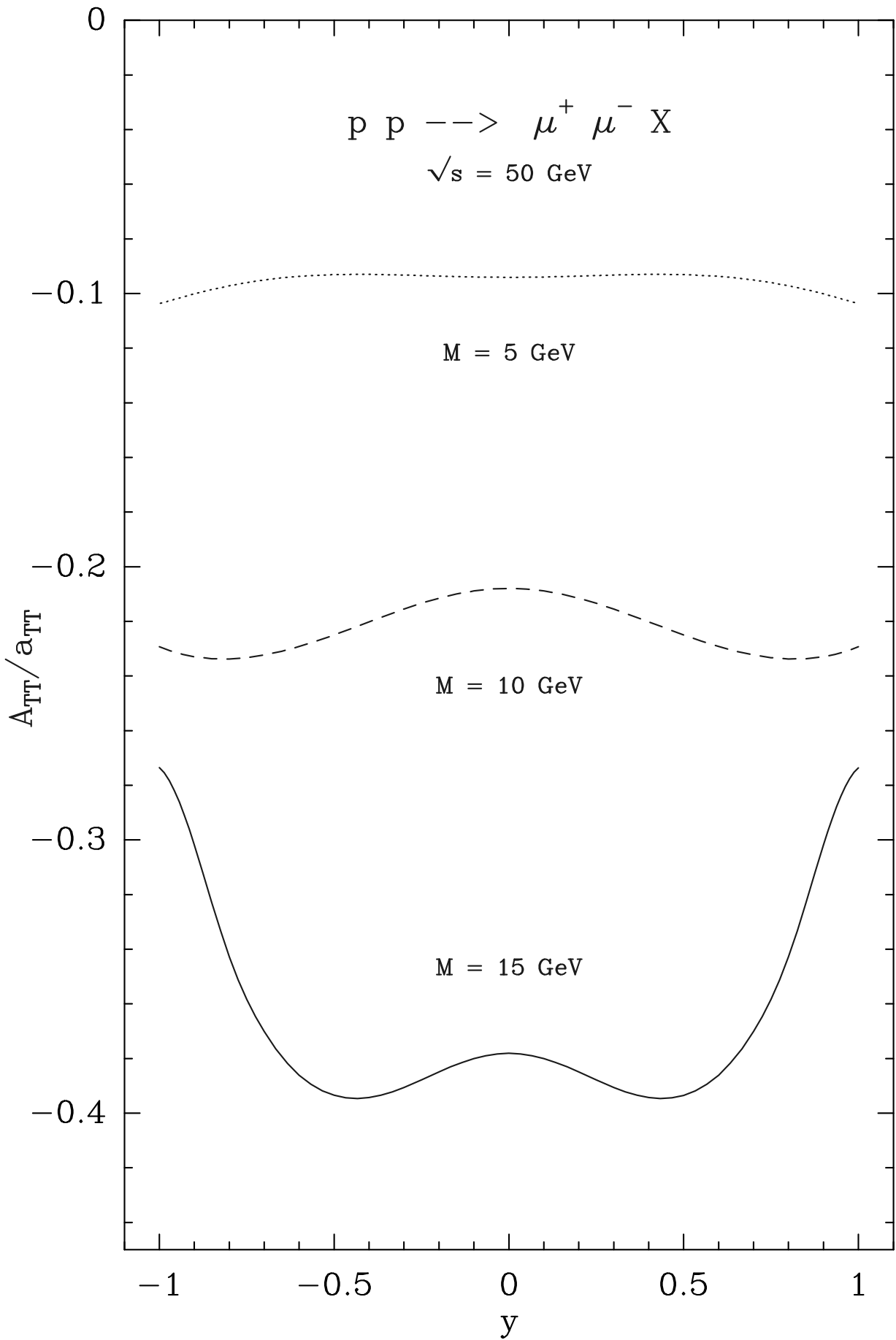
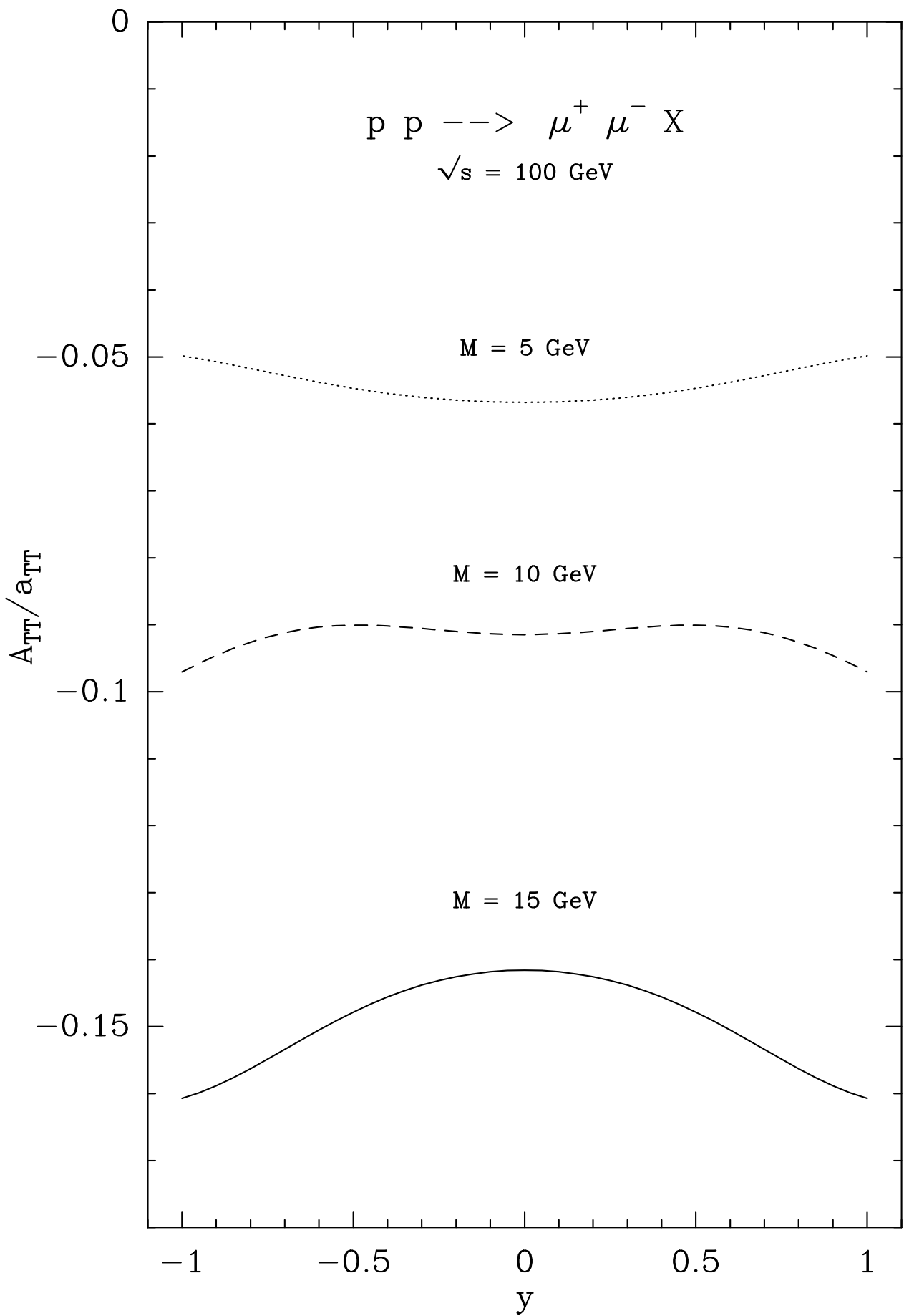
\end{document}